\documentclass[amsmath,amssymb]{revtex4-1}

\usepackage{amscd, graphicx}
\usepackage[all]{xy}

\usepackage{color}
\usepackage{comment}
\newcommand{\ins}[1]{\textcolor{black}{#1}}
\newcommand{\revs}[1]{\textcolor{black}{#1}}

\makeatletter
\@addtoreset{equation}{section}
\makeatother
\usepackage{enumerate}

\def\ii{{\mathrm{i}}}

\def\hht{{\hat{t}}}
\def\hhr{{\hat{r}}}

\def\ttt{{\tilde{t}}}
\def\ttr{{\tilde{r}}}

\def\ee{\mathrm e}

\def\cN{{\mathcal{N}}}
\def\cE{{\mathcal{E}}}
\def\cS{{\mathcal{S}}}
\def\cU{{\mathcal{U}}}

\def\CC{{\mathbb C}}
\def\ZZ{{\mathbb Z}}

\def\RR{{\mathbb R}}

\def\cF{{\mathcal{F}}}

\def\UU{{\mathrm{U}}}

\def\book#1{\rm{#1}, }
\def\paper#1{\textit{#1}, }
\def\jour#1{\rm{#1}, }
\def\yr#1{({\rm{#1}) }}
\def\vol#1{\textbf{#1}}
\def\pages#1{\rm{#1}}

\def\publaddr#1{\rm{#1}, }
\def\publ#1{\rm{#1}, }
\def\by#1{{\rm{#1}, }}

\begin{document}


\title{New theory of diffusive and coherent nature of optical wave via a quantum walk}

\author{Yusuke Ide}
 \affiliation{
Department of Information Systems Creation, 
Faculty of Engineering, Kanagawa University
Kanagawa, Yokohama 221-8686, Japan}
\email{ide@kanagawa-u.ac.jp}

\author{Norio Konno}
 \affiliation{
Department of Applied Mathematics, 
Faculty of Engineering, Yokohama National University
Hodogaya, Yokohama 240-8501, Japan}
\email{konno@ynu.ac.jp}

\author{Shigeki Matsutani}
 \affiliation{
Industrial Mathematics, National Institute of Technology, 
Sasebo College, 1-1, Okishin-machi, Sasebo, Nagasaki 857-1193, Japan}
\affiliation{
Institute of Mathematics for Industry, Kyushu University, 
Motooka 744, Nishi-ku, Fukuoka 819-0395, Japan}
 \email{smatsu@sasebo.ac.jp}

\author{Hideo Mitsuhashi}
 \affiliation{
Faculty of Education, Utsunomiya University
Utsunomiya, Tochigi 321-8505, Japan}
 \email{mitsu@cc.utsunomiya-u.ac.jp}

\date{\today}

\begin{abstract}
We propose a new theory on a relation between diffusive and coherent nature 
in one dimensional wave mechanics
based on a quantum walk.
It is known that 
the quantum walk in homogeneous matrices
provides the coherent property of wave mechanics.
Using the recent result of a localization phenomenon
in a one-dimensional quantum walk 
(Konno {\it{Quantum Inf. Proc.}} \yr{2010} \vol{9} \pages{405-418}),
we numerically show that the randomized 
localized matrices suppress the coherence and give diffusive nature.
\end{abstract}

\maketitle

\section{Introduction}

Feynman presented the nature of wave mechanics based on his picture of 
path integral \cite{FH}. 
At a point in the space
the value of a wave function which is a complex number
is determined by
the summation of contributions over all paths from all other
point in the space \cite{FH}.
In his lecture for an ordinary audience \cite{F}, 
he explained the 
behavior of the light ray stemmed from a point to another point
along the line of his picture
and showed that  
the picture reproduces the
Fermat principle of the light.
Though Feynman mentioned
nature of the quantum mechanics,
the natural correspondence between quantum mechanics and
optics as wave mechanics \cite{BW}
means that nature of light is coherence
and thus the interference naturally appears in optics.
A path from a point to another point, in one-dimensional case,
consists of transmission and reflection at each point, and 
\revs{
for each path, a
}
complex value is determined. 
By summing the complex values over all 
\revs{
possible 
}
paths, the value of the wave function at the point is determined.

In optics, the \revs{
transmission and reflection
} in a one-dimensional
system is described
\revs{by $S$-matrix in the transfer matrix theory \cite{C,L}.
}
\ins{
Crook proved that the transfer matrix theory and path integral
(summation) are equivalent statically \cite{C}.
}

Recently such a picture is obtained in quantum walks mathematically.
The reviews for developments of the theory of quantum walks are, 
for example, Kempe \cite{Kem}, Kendon \cite{Ken}, 
Venegas-Andraca \cite{VAndraca2008, VA}, Konno \cite{Konno2008b}, 
Manouchehri and Wang \cite{ManouchehriWang2013}, Portugal \cite{Portugal2013}. 
Konno analytically showed that
the homogeneous scattering by Hadamard type 
matrix
at each point
 forms the {\lq\lq}wavefront{\rq\rq} in 
quantum walks \cite{K05,Konno2008b}. 
It could be interpreted as a rigorous mathematical proof of
the picture in Feynman's lecture \cite{F}.
Quantum walks provide the nature of the wave mechanics,
in which the complex number plays a crucial role.
As the probability density  corresponds to the intensity of light,
the results of quantum walks exhibit the coherent nature,
the existence of the wavefront, and 
the Fermat principle of light in a one-dimensional system
 as in Sec.\ref{sec:Optics}.
\ins{
Further as the optical ray has the finite speed,
quantum walks naturally include the time-development and  
have the finite speed.}

On the other hand, 
when we consider coloring of the colored materials, e.g.,
paint, ink in a media, colored fabric and so on,
the color is determined by the diffusive reflection of the light
with 
\revs{
the intensity distribution of 
the visible wavelengths.
}
The incident white rays with flat spectrum distribution penetrates
into a colored material and then some of its spectrum
are absorbed and scattered. We observe the out-going diffusive
reflection colored rays whose distribution of the intensity on
the visible wavelengths differs from white one. 
The Kubelka-Munk model which is a one-dimensional
model describes well the nature of coloring of diffusive reflection
of the optical rays except \revs{structural} color.
We decompose the white ray to each monochromatic ray of each
wavelength and
the behavior of each monochromatic ray is governed by 
a differential equation \cite{KM} (See Appendix B).
In the equation, we handle the light intensity of a fixed
spectrum as a real number rather than a complex number at each point.
In other words, in the model, coherence and interference of
the light are neglected though coherence is the most important
nature of the light as mentioned above.
It implies that
in the coloring, 
the real valued intensity plays much more 
important roles than complex valued properties of the coherence.
The real valued ray is based on the diffusive light theory \cite{D,KM}.
Light is scattered by small particles and voids in the media
many times, e.g., following the Mie scattering theory
and absorbed by colored materials. It is said that
these scattering and absorption make the coherent light forget its 
coherence and become the diffusive light.

It is a natural problem when the optical wave forgets its coherence.
We numerically consider this problem using quantum walks
and the correspondence
between quantum mechanics and optics as in Sec.\ref{sec:Optics}.

\revs{
One of the answers was obtained by Ribeiro, Milman, and Mosseri \cite{RMM},
Mackay, Bartlett, Stephenson and Sanders \cite{MBSS} and
Kempe \cite{Kem}.}
They studied the quantum random walk which is a
 model of quantum walks with randomized
quantum-coin matrices.
The randomized quantum-coin matrices cause quantum walks
to lose its coherence of wave
nature and we have the diffusive and localized distributions.
However they assumed that the matrices change their value randomly
in space and time.
The assumption differs from the above situation.
Small particles, voids and colored materials in a solid
are static.  

\ins{
Another answer was obtained by Haney and van Wijk \cite{HvW}.
They studied the wave transport in  the random media 
and revised the Kubelka-Munk equation (they also called it
O'Doherty-Anstey formula \cite{HvWS}) based on the theoretical
investigation of Goedecke \cite{G} and the numerical model proposed
by Weaver and his coauthors \cite{LW,W1,W2,WB}.
The numerical model is a coupled spring model with randomized
spring constants and reproduces many interesting phenomenon,
such as Anderson's localization phenomenologically \cite{LW,W1,W2,WB}.
Since 
Goedecke showed that
under the assumption that randomness makes the wave lose its coherency,
the transfer matrix theory is connected with the Kubelka-Munk theory,
Haney and van Wijk used the transfer matrix theory phenomenologically
and derived a time-development equation as
a revision of the formula in the Kubelka-Munk theory.
They analyzed their numerical computations of the 
random coupled spring model and showed that the equation reproduces
the numerical results.
}

\ins{
Though the numerical results are similar to ours in this article,
the problem is not solved in the framework of the path integral 
(path summation) theory and the transfer matrix theory
\cite{C,L}.
}

\ins{
Our purpose of this article is to solve  the problem when
the wave forget its coherency from the first principle of
the path integral (path summation) and the transfer matrix theory.
}

\ins{
For the purpose, we should avoid
numerical errors nor other numerical problems.
For example, the logistic map whose solutions are known as chaotic
for certain parameters could be regarded as a discretization
of the logistic equation, and 
a discrete version of the logistic equation
depending on its discretization method sometimes loses
the chaotic behavior \cite{YS}. 
Further it is well-known that 
the numerical diffusion of the wave function
is a crucial problem in computational fluid dynamics \cite{M}.
In other words, 
discretization of a continuum system sometimes
shows non-trivial aspects.
}

\ins{
The quantum walk is a very simple model and has been 
rigorously investigated as mentioned above.
\cite{Kem,Ken,VAndraca2008,VA,Konno2008b,ManouchehriWang2013,Portugal2013}. 
In fact, using quantum walks, many physical
phenomenon were re-investigated and reevaluated rigorously.
}

\ins{
Using the correspondence between
optical transfer matrix theory and quantum walks,
we naturally introduced the time development
(see Appendix A).
Thus we will investigate the problem in the framework of quantum
walks rigorously.
}

Recently Konno gave an interesting fact on the localization in
quantum walks \cite{K10} that a single 
\revs{quantum coin} 
matrix causes
the localization of the density.
We call the matrix B-impurity as in Sec \ref{sec:QW}.

In this article, we employ this model and consider the 
\revs{quantum coin} 
matrices which consist of 
the random B-impurities in the homogeneous material.
We set sufficiently many B-impurities uniformly randomly
in homogeneous Hadamard matrices as a background,
whose density $p = 0.05$ in 6000 points.
Mathematically it means the Poisson point process of the localized matrix
of the B-impurity in the homogeneous material.
We fix these 
\revs{quantum coin} 
matrices and consider quantum walks over them.
These impurities are static like the voids, and
 the small particles in the solid.
It implies one of  disordered models of quantum walk
\cite{K05b,RMM,MBSS,Kem}.
There we numerically show that such impurities suppress the coherent nature
and the distribution of quantum walks behaves like diffusive one.

In other words, we provide a new phenomena of the connection
between the coherency and diffusive nature of optical wave
using quantum walks.

The contents in this article is as follows:
Sec \ref{sec:QW} gives a brief review of the quantum walks
including the recent results of localization, 
which we call A- and B- impurities,  \cite{K10}.
In Sec \ref{sec:Optics},
we interpret the models and results in quantum walks as
optical systems.
Sec \ref{sec:PPP} gives the numerical results to answer 
the problem mentioned above.
We discuss our results in Sec \ref{sec:Dis}.
We also give the short reviews of
the correspondences among transfer matrix theory, path integral
(path summation), and quantum walks in 
in \ref{sec:TMPIQW}, and 
the Kubelka-Munk model  in \ref{sec:KMT}.

\section{Short review of Quantum Walks}
\label{sec:QW}

Following \cite{Konno2008b,VA}, we show a short review of quantum walks in 
this section.
\revs{
Let us consider the infinite one-dimensional lattice
$$
\cN=\{ n a \ | \ n \in \ZZ\}, \quad
\cE=\{ [n a, (n+1)a] \ | \ n \in \ZZ\},
$$
and the set 
of the complex two dimensional vector valued functions 
of $\cN$,
denoted by $\cS(\cN):=\{\sigma\}$,
}
$$
     \sigma= \left\{\sigma_n:=\begin{pmatrix} \sigma^{(+)}_n\\ 
                                              \sigma^{(-)}_n
                       \end{pmatrix} \ \Bigr| \sigma_{\pm, n}\in \CC, 
                   n \in \ZZ\right\}.
$$
\ins{
Here $\ZZ$, $\RR$ and $\CC$ are the sets of integers, real numbers,
and complex numbers respectively.
}
Further we consider a subset $\cS_0(\cN)$ of
$\cS(\cN)$,
$$
     \cS_0(\cN) = \left\{\sigma \in \cS(\cN) \ \Bigr| \ 
                 |\sigma|:=\sum_i 
(|\sigma^{(+)}_{i}|^2 +|\sigma^{(-)}_{i}|^2) = 1\right\},
$$
and $p:\cS_0(\cN) \to \cF_0(\cN)$
($p(\sigma)=\{|\sigma^{(+)}_{i}|^2 +|\sigma^{(-)}_{i}|^2\}_i$),
where
\revs{
$$
     \cF_0(\cN) = \left\{d_n \in \RR \Bigr| \ n \in \ZZ, d_n\ge 0,
\sum_{n\in \ZZ} d_n = 1 \right\}.
$$
}
An element $\sigma$ of $\cS_0(\cN)$ means a quantum state in quantum
walks
\revs{
and $p(\sigma)$ means the probability density of the state $\sigma$. 
}
Later we consider $p(\sigma)$ as the intensity of the light 
using the correspondence between quantum mechanics and optics.
We consider a  linear map from 
$\cS_0(\cN)$ to $\cS_0(\cN)$ which is given by
$$
     \tau = \left\{\tau_n:=\begin{pmatrix}
     \tau^{(++)}_{n} & \tau^{(+-)}_{n} \\
     \tau^{(-+)}_{n} & \tau^{(--)}_{n} \end{pmatrix} \ \Bigr| 
     \ \tau_{n} \in \UU(2), \    n \in \ZZ\right\}
$$
with its action $\tau \sigma$, 
\revs{
\begin{equation}
(\tau \sigma)_n = \begin{pmatrix}
 \tau^{(++)}_{n-1} \sigma^{(+)}_{n-1} +\tau^{(+-)}_{n-1} \sigma^{(-)}_{n-1} \\
 \tau^{(-+)}_{n+1} \sigma^{(+)}_{n+1} +\tau^{(--)}_{n+1} \sigma^{(-)}_{n+1} \\
\end{pmatrix}.
\label{eq:tausigma}
\end{equation}
}
The set of the linear maps is denoted by $\cU(\cN)$.
\revs{
$\UU(2)$ means the set of unitary matrix, i.e., 
for every $A \in \UU(2)$,
 its hermite conjugate $A^*$ is identified with its
inverse matrix. The unitary matrix is referred the 
quantum-coin matrix
and the element $\cU(\cN)$ is the 
quantum-coin operator.
}

\begin{figure}[ht]
 \begin{center}
\includegraphics[width=7cm]{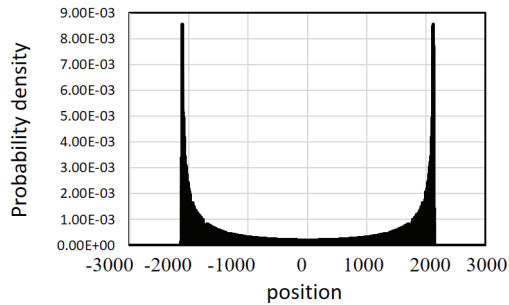}
 \newline
 \end{center}
\caption{
The Probability density of the Hadamard process 
at $t=3000$.}
\label{fig:HP}
\end{figure}

When an element $\tau \in \cU(\cN)$ 
is homogeneous case or the constant for every $n$,
the behavior is completely determined by Konno \cite{K05,Konno2008b}.
When every $\tau_n$ of $\tau\in \cU(\cN)$
is given by
$$
\frac{1}{\sqrt{2}}
\begin{pmatrix} 1 & 1 \\ 1 & -1 \end{pmatrix},
$$
which is called Hadamard type.
We let it be denoted by $\tau^H\in \cU(\cN)$.
\revs{
The probability density $p(\sigma)$ for the initial
state,
\begin{equation}
\sigma_n = \left\{ \begin{matrix}
\frac{1}{\sqrt{2}}\begin{pmatrix}1\\1\end{pmatrix},& n=0,\\
\begin{pmatrix}0\\0\end{pmatrix},&\mbox{otherwise}, 
\end{matrix}\right.
\label{eq:II2}
\end{equation}
is illustrated in Figure \ref{fig:HP} at $t=3000$. 
For this case, the limit behavior of
the probability density is given by \cite{K05,Konno2008b},
}
\begin{equation}
\lim_{t\to \infty}
 P(u\le \frac{X_t}{t}\le v) = \int^v_u 
\frac{ 1}
{\pi \sqrt{1-x^2}\sqrt{1-2x^2}} d x,
\label{eq:1}
\end{equation}
if the initial state $t=0$ is emitted from the point $n=0$,

As mentioned in Introduction, the probability density 
has the wavefront and 
recovers the dispersion relation of the light ray, which 
is consistent with the picture of Feynman \cite{F,FH}.
In other words, the peaks in the envelop 
of (\ref{eq:1}) and Figure \ref{fig:HP}.
represent
the wavefront of the ballistic motion
\ins{ for the initial state  
(\ref{eq:II2}).}
Using the correspondence between quantum mechanics and optics,
it is interpreted as the wavefront of the ballistic motion 
of light whose speed is $1/\sqrt{2}$.
The probability density is regarded as the intensity of  
light ray as in Sec. \ref{sec:Optics}
and \ref{sec:TMPIQW}.

\begin{figure}[ht]
 \begin{center}
\includegraphics[width=6cm]{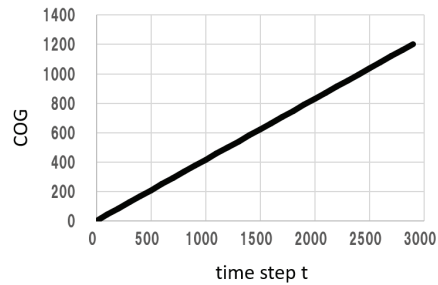}
 \newline
 \end{center}
\caption{
The time dependence of the center of gravity of the half side of
the density distribution of Hadamard process.}
\label{fig:COGHard}
\end{figure}

The probability density is symmetric with respect to the origin
\ins{
as in Figure \ref{fig:HP},
were we set the origin as the center of the system.
}
We introduce the center of gravity of the half-side,
\begin{equation}
 \mathrm{COG}(t) := 
\frac{\sum_{n \ge 0} n\sigma_n^* \sigma_n}
{\sum_{n \ge 0} \sigma_n^* \sigma_n}.
\label{eq:1.1}
\end{equation}
We numerically compute the time dependence of the center of gravity
of the half-side, which is displayed in Figure \ref{fig:COGHard}.
Then the numerical computation shows that its value is given by 
$$
 \mathrm{COG}(t) = \beta_0 t^{\alpha_0}
$$
where $\alpha_0=1$.
It is proved strictly using the 
equation (\ref{eq:1}).

\subsection{Localization in Quantum Walk}

Recently Konno also investigated the behaviors of
quantum walks for the configurations of $\tau^A$ and $\tau^B\in \cU(\cN)$:
$$
\tau_n^A = 
\left\{
\begin{matrix}
\frac{1}{\sqrt{2}}
\begin{pmatrix} \ee^{\ii \gamma} & 1 \\ 1 & -\ee^{-\ii \gamma} \end{pmatrix},
& \mbox{for }n = 0,\\
\frac{1}{\sqrt{2}}
\begin{pmatrix} 1 & 1 \\ 1 & -1 \end{pmatrix},
& \mbox{otherwise},\\
\end{matrix}\right.
$$
$$
\tau_n^B = 
\left\{
\begin{matrix}
\frac{1}{\sqrt{2}}
\begin{pmatrix} 1 & \ee^{\ii \gamma} \\ \ee^{-\ii \gamma} & -1 \end{pmatrix}.
& \mbox{for }n = 0,\\
\frac{1}{\sqrt{2}}
\begin{pmatrix} 1 & 1 \\ 1 & -1 \end{pmatrix},
& \mbox{otherwise}.\\
\end{matrix}\right.
$$
In this article, we call the former case A-type \cite{K09} 
and the latter one B-type \cite{K10}.
Konno proved that A-type does not cause the localization whereas B-type does
\cite{K09,K10}.
Further we also call the matrix $\tau^A_0$ A-impurity
and the matrix $\tau^B_0$ B-impurity here.

\begin{figure}[ht]
\begin{minipage}{0.4\hsize}
 \begin{center}
\includegraphics[width=6cm]{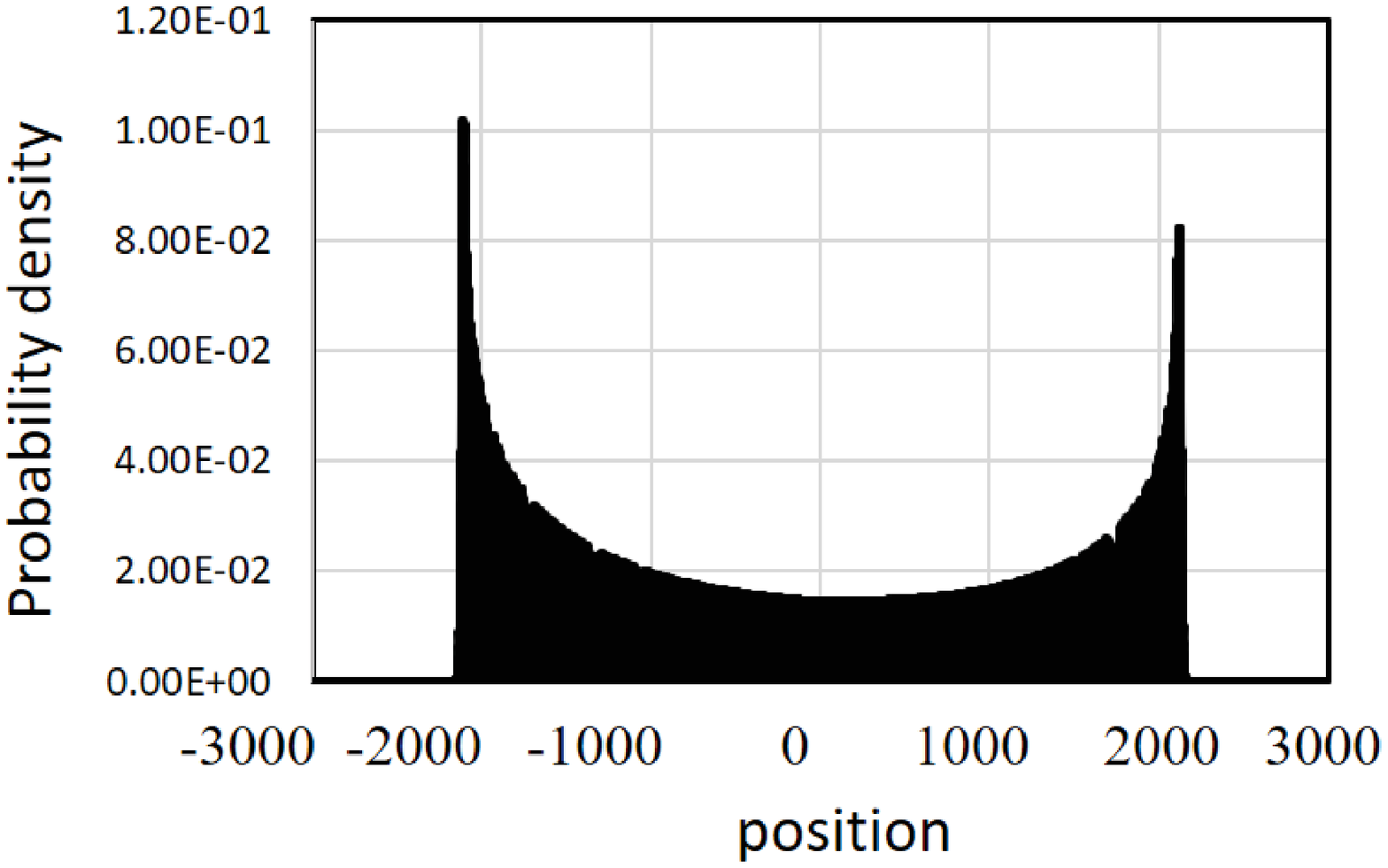}
 \newline
\bigskip 
(a)
 \end{center}
\end{minipage}
\begin{minipage}{0.4\hsize}
 \begin{center}
\includegraphics[width=6cm]{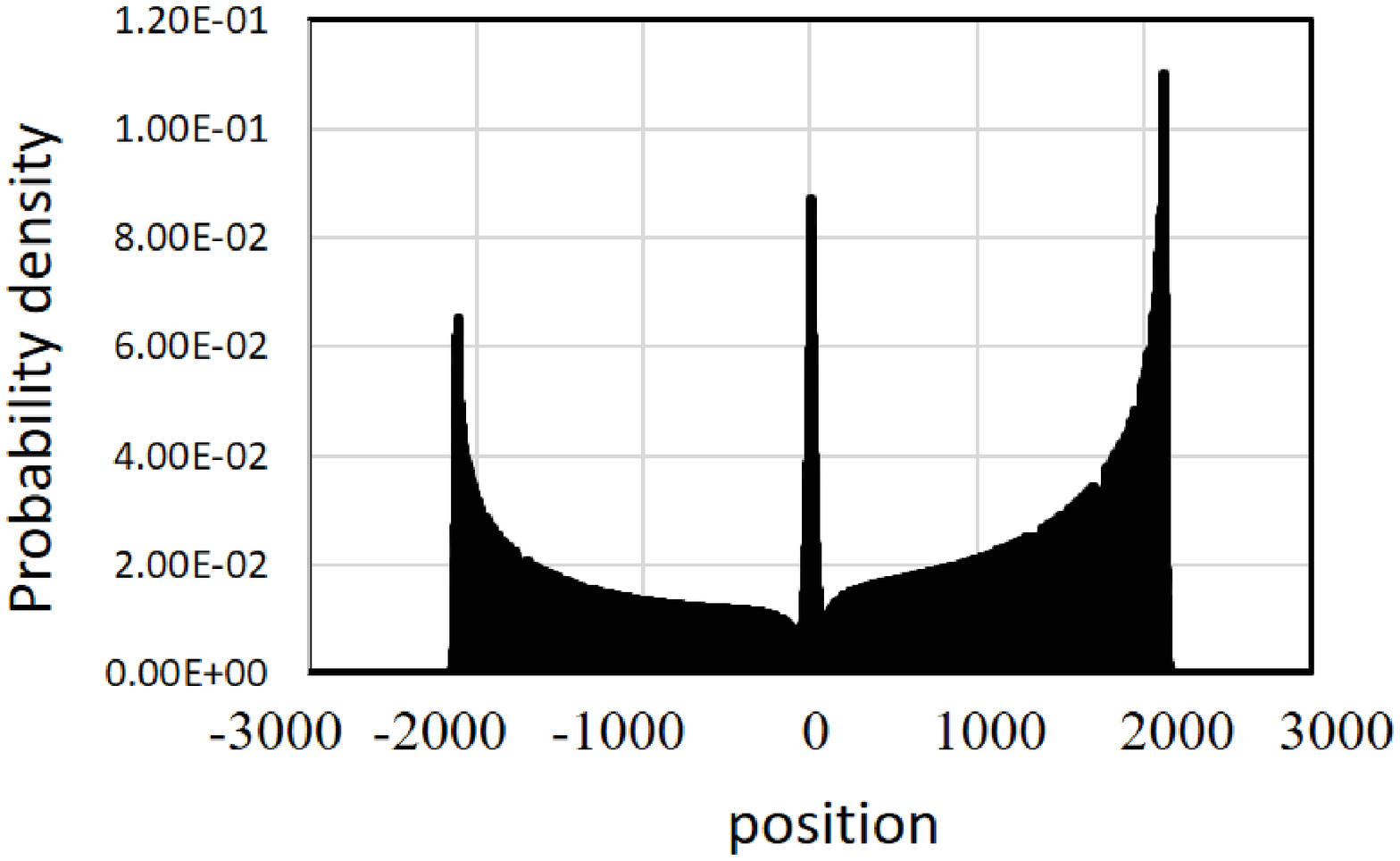}
 \newline
\bigskip 
(b)
 \end{center}
\end{minipage}
\begin{minipage}{0.4\hsize}
 \begin{center}
\includegraphics[width=6cm]{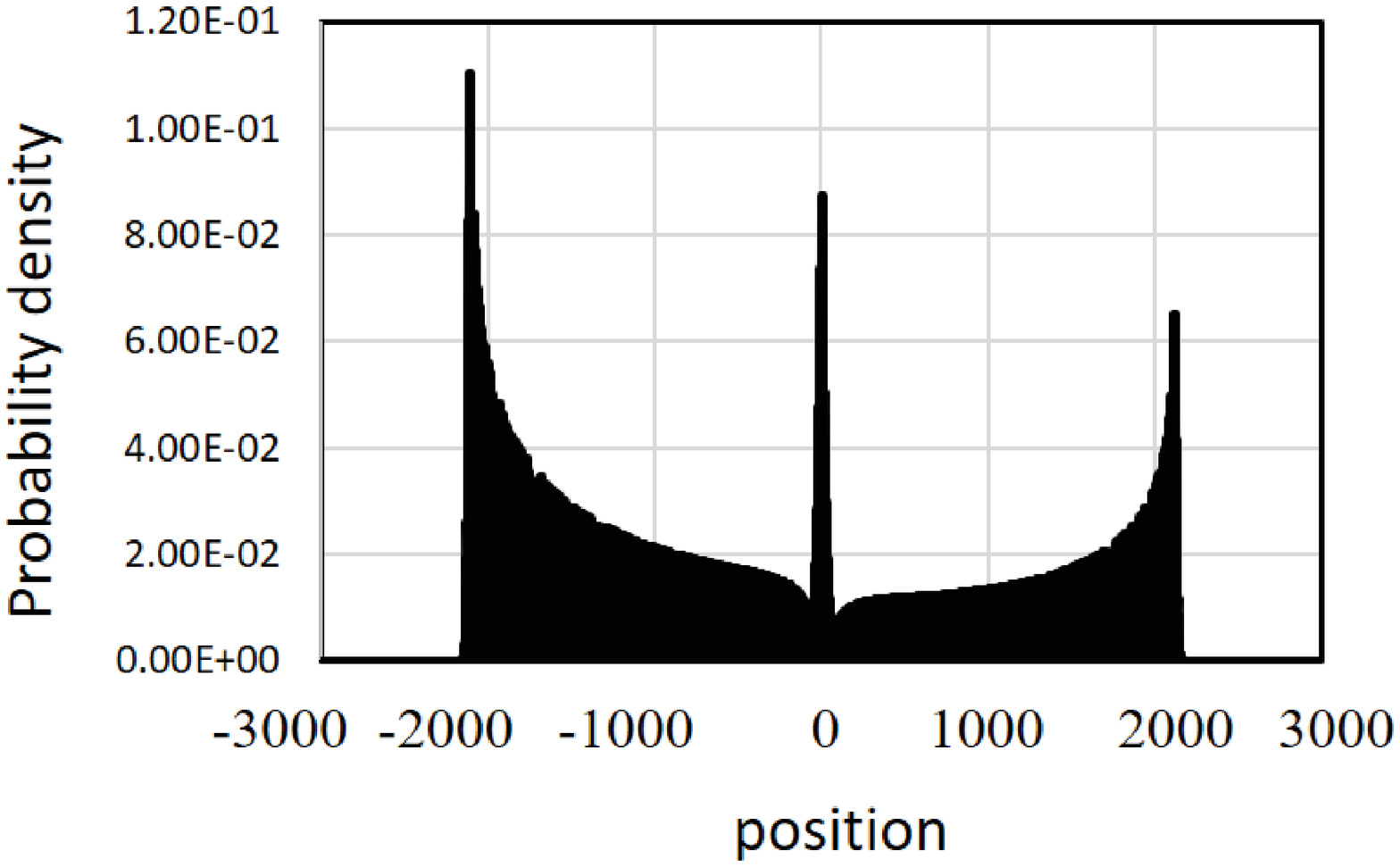}
 \newline
\bigskip 
(c)
 \end{center}
\end{minipage}
\caption{
\revs{
The Probability densities of the 
A-type and B-type processes:
(a) is the the A-type process of $\gamma=0.3$ at $t=3000$,
and (b) and (c) are the B-type processes  of $\gamma=0.3$ and 
$\gamma=-0.3$ at $t=3000$ respectively.
}
}
\label{fig:ABtype}
\end{figure}

We compute these quantum walks for 6000 points
numerically.
The numerical computations reproduce the results of Konno
on the probability densities of A-type and 
B-type matrices.
We computed $\gamma = 0.3$ cases of A-type in 
\revs{Figure \ref{fig:ABtype}} (a)
and B-type in Figure \ref{fig:ABtype} (b) for
the initial state (\ref{eq:II2}).
The probability densities are plotted there.
In the A-type case, we cannot observe the localization of the 
probability densities whereas in the B-type case, we find
the localization.

\begin{table}[httb]
\begin{center}
\caption{Table: $2^{(t+1)/2}\sigma_n$ A-impurity $\delta=\ee^{\ii \gamma}$ }
  \begin{tabular}{|l|c|c|c|c|c|c|}
\hline
$t\backslash$ &  -2 & -1 & 0 & 1 & 2 & 3 \\
\hline
0 &
$\begin{pmatrix} 0\\0 \end{pmatrix}$ & 
$\begin{pmatrix} 0\\0 \end{pmatrix}$ & 
$\begin{pmatrix} 1\\ 1 \end{pmatrix}$ & 
$\begin{pmatrix} 0\\0 \end{pmatrix}$ & 
$\begin{pmatrix} 0\\0 \end{pmatrix}$ & 
$\begin{pmatrix} 0\\0 \end{pmatrix}$  
\\
\hline
1 &
$\begin{pmatrix} 0\\0 \end{pmatrix}$ & 
$\begin{pmatrix} 0 \\ 1-\frac{1}{\delta} \end{pmatrix}$ & 
$\begin{pmatrix} 0\\0 \end{pmatrix}$ & 
$\begin{pmatrix} \delta +1\\ 0 \end{pmatrix}$ & 
$\begin{pmatrix} 0\\0 \end{pmatrix}$ & 
$\begin{pmatrix} 0\\0 \end{pmatrix}$  
\\
\hline
2 &
$\begin{pmatrix} 0 \\ -1+\frac{1}{\delta} 
\end{pmatrix}$ & 
$\begin{pmatrix} 0\\0 \end{pmatrix}$ & 
$\begin{pmatrix} 
1-\frac{1}{\delta}\\\delta+1\end{pmatrix}$ & 
$\begin{pmatrix} 0\\0 \end{pmatrix}$ & 
$\begin{pmatrix} \delta+1 \\ 0 \end{pmatrix}$ & 
$\begin{pmatrix} 0\\0 \end{pmatrix}$  
\\
\hline
3 &
$\begin{pmatrix} 0\\0 \end{pmatrix}$ & 
$\begin{pmatrix} -1+\frac{1}{\delta} \\
 -\frac{2}{\delta}  
\end{pmatrix}$ & 
$\begin{pmatrix} 0\\0 \end{pmatrix}$ & 
$\begin{pmatrix} 2\delta \\  \delta+1 
\end{pmatrix}$ & 
$\begin{pmatrix} 0\\0 \end{pmatrix}$ & 
$\begin{pmatrix} 1+ \delta \\ 0 \end{pmatrix}$  
\\
\hline
  \end{tabular}
\end{center}
\end{table}

\begin{table}[htb]
\begin{center}
\caption{Table: $2^{(t+1)/2}\sigma_n$ B-impurity  $\delta=\ee^{\ii \gamma}$}
  \begin{tabular}{|l|c|c|c|c|}
\hline
$t\backslash$ & -1 & 0 & 1 & 2  \\
\hline
0 &
$\begin{pmatrix} 0\\0 \end{pmatrix}$ & 
$\begin{pmatrix} 1\\ 1 \end{pmatrix}$ & 
$\begin{pmatrix} 0\\0 \end{pmatrix}$ & 
$\begin{pmatrix} 0\\0 \end{pmatrix}$  
\\
\hline
1 &
$\begin{pmatrix} 0 \\ \frac{1}{\delta}-1 \end{pmatrix}$ & 
$\begin{pmatrix} 0\\0 \end{pmatrix}$ & 
$\begin{pmatrix} 1+ \delta \\ 0 \end{pmatrix}$ & 
$\begin{pmatrix} 0\\0 \end{pmatrix}$  
\\
\hline
2 &
$\begin{pmatrix} 0\\0 \end{pmatrix}$ & 
$\begin{pmatrix} 
  \frac{1}{\delta}-1\\1+ \delta\end{pmatrix}$ & 
$\begin{pmatrix} 0\\0 \end{pmatrix}$ & 
$\begin{pmatrix} 1+ \delta \\ 0 \end{pmatrix}$  
\\
\hline
3 &
$\begin{pmatrix} -\frac{1}{\delta}+1 \\
 \frac{1}{\delta^2}-\frac{1}{\delta} -1- \delta 
\end{pmatrix}$ & 
$\begin{pmatrix} 0\\0 \end{pmatrix}$ & 
$\begin{pmatrix} 
\frac{1}{\delta}-1+\delta + \delta^2 \\ 1+ \delta \end{pmatrix}$ & 
$\begin{pmatrix} 0\\0 \end{pmatrix}$ 
\\
\hline
  \end{tabular}
\end{center}
\end{table}

The asymmetry in \revs{Figure \ref{fig:ABtype} (b)} comes from the
phase $\gamma$; if we put $\gamma = -0.3$, we obtained 
its reflection image as in \revs{Figure \ref{fig:ABtype} (c)}.
We show explicit states around $n=0$ of a few time steps
in Tables 1 and 2.
In \revs{Figure \ref{fig:ABtype}} ,
 we also observe the coherent wave, 
\ins{
which is ballistic like Figure \ref{fig:HP},
}
besides the localized
one which is localized in the center.
It is natural because the wave must obey the superposition
principle.

\begin{figure}[ht]
\begin{minipage}{0.33\hsize}
 \begin{center}
\includegraphics[width=5cm]{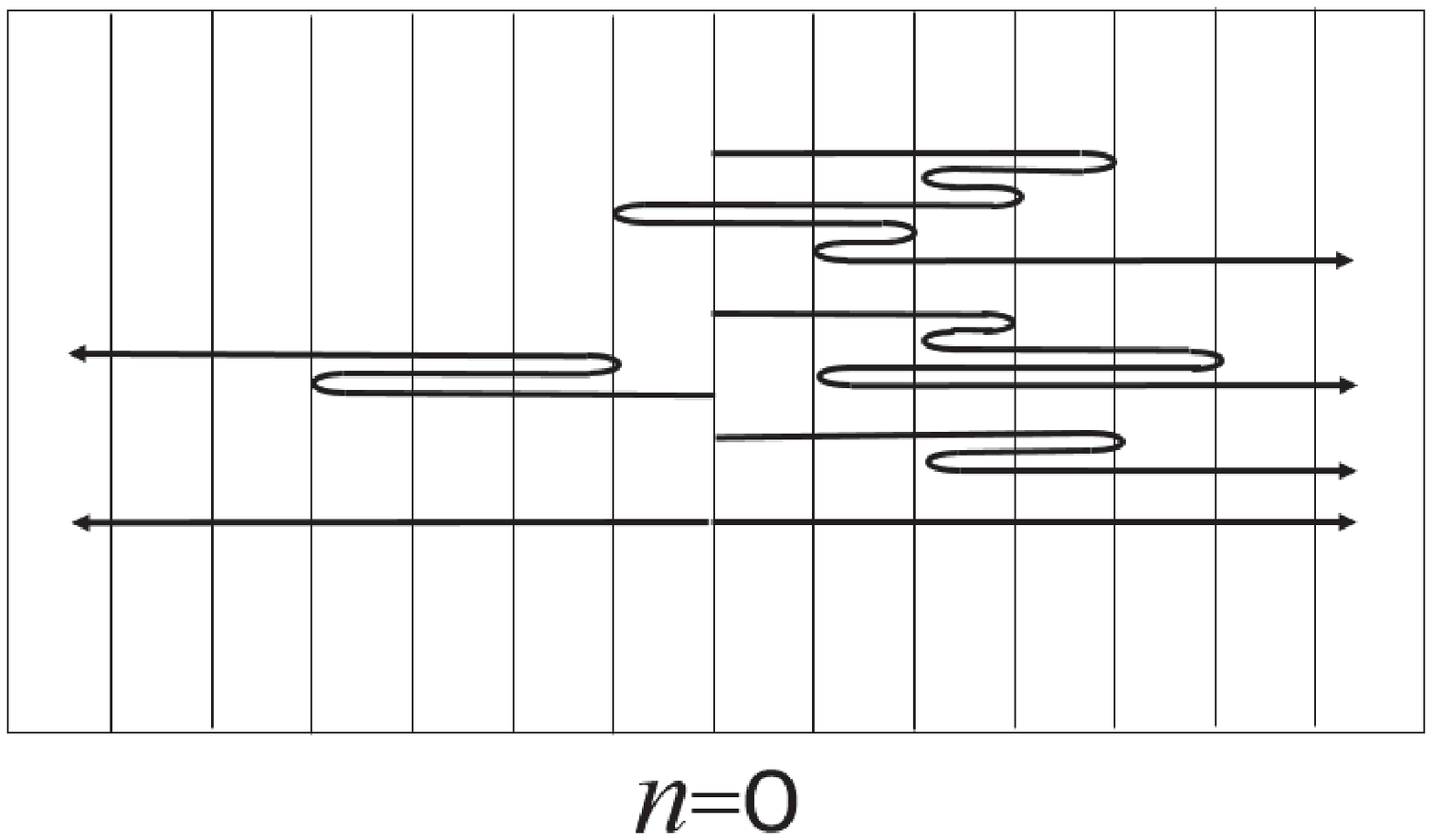}
 \newline
(a)
 \end{center}
\end{minipage}
\begin{minipage}{0.33\hsize}
 \begin{center}
\includegraphics[width=5cm]{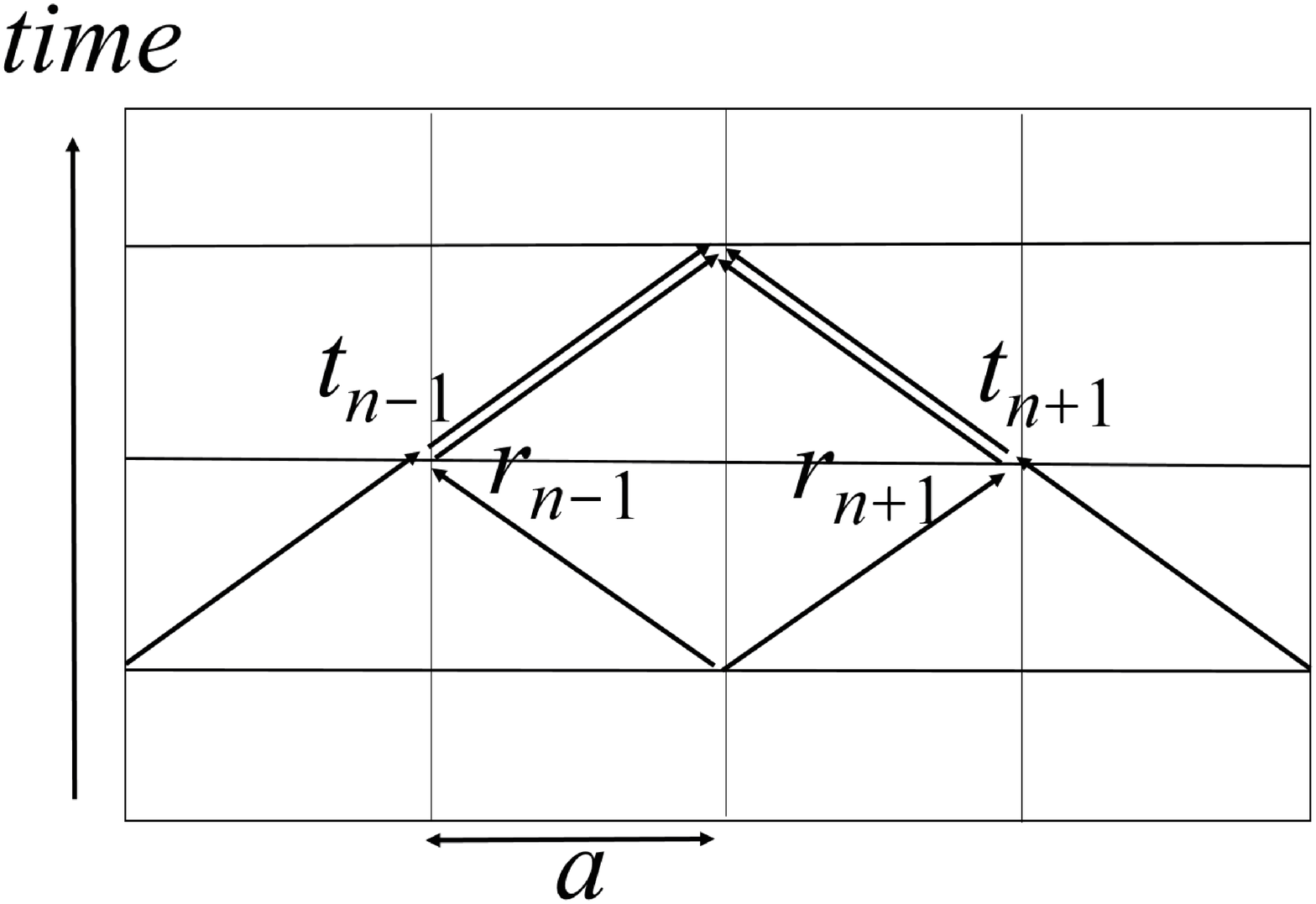}
 \newline
(b)
 \end{center}
\end{minipage}
\begin{minipage}{0.33\hsize}
 \begin{center}
\includegraphics[width=5cm]{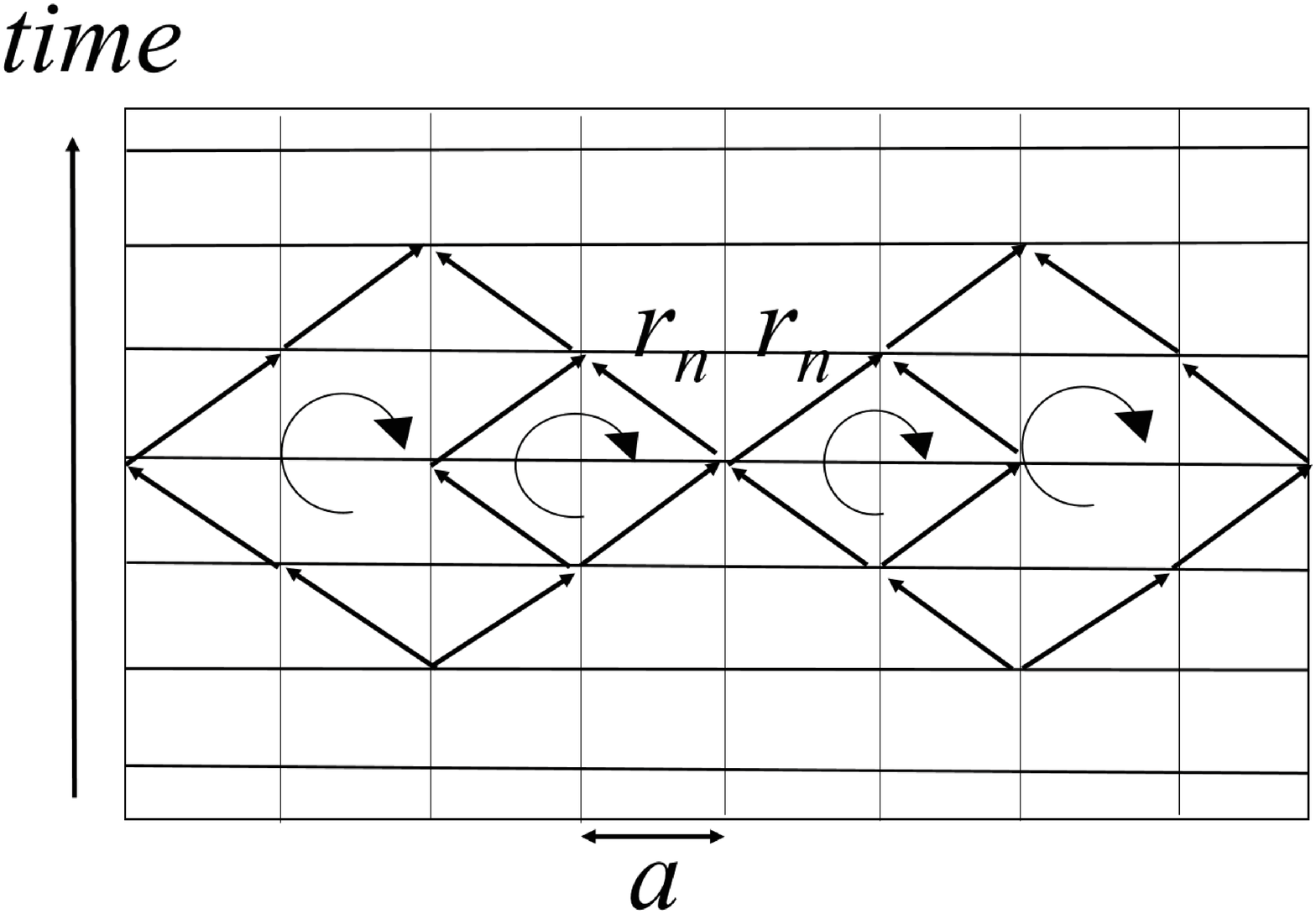}
 \newline
(c)
 \end{center}
\end{minipage}
\caption{
Optical Layer}
\label{fig:OL}
\end{figure}

\bigskip


\bigskip 
\bigskip

\section{Quantum Walk from Optical point of view}
\label{sec:Optics}

In this section,
let us consider the phenomena in quantum walks
from the viewpoint of optics.
They correspond to ideal optical layers
whose interfaces agree with $\cN$
as in Figure \ref{fig:OL} (a) \cite{BW}.
In the layers, the light is reflected
at the interface with a complex reflection rate $\ttr$
or transmits to next layer with transmission 
\revs{rate} $\ttt$.

It implies that we regard a quantum-coin matrix in quantum walks
as a 
 \revs{
$S$-matrix in the scalar optics (see \ref{sec:TMPIQW})}.
We consider an optical ray lit at the position $n=0$ and time $t=0$.
By letting its wavevector be 
\revs{
$k_n$ for $n$-th segment in $\cE$.
} the phase is shifted
to $\ee^{\ii k_n a}$ when it passes through a layer.
The finite speed of the light restricts the possible paths
among the interfaces. 
The amplitude of the light at the $m$-step of time direction 
at $n$-th interface is denoted by $\sigma_{n,m}$.
Since the complex amplitude of the light is given by the two states,
i.e., the left-direction and the right-direction
\revs{
(see \ref{sec:TMPIQW}),
\begin{gather*}
\begin{split}
\psi(x,t=m) &= 
\sigma^{(+)}_{n,m}\ee^{\ii k_{n-1}(x- n a)}
+ \sigma^{(-)}_{n,m}\ee^{-\ii k_n(x- n a)},
\end{split}
\end{gather*}
the time evolution $\tau \in \cU(\cN)$ is regarded as
\begin{gather}
\begin{split}
\sigma^{(+)}_{n,m} &=
\ttt_{n-1} \sigma^{(+)}_{n-1,m-1} 
+\ttr_{n-1} \sigma^{(-)}_{n-1,m-1} \\
\sigma^{(-)}_{n,m} &=
\ttr_{n+1}' \sigma^{(+)}_{n+1,m-1} 
+\ttt_{n+1}'\sigma^{(-)}_{n+1,m-1}. \\
\end{split}
\label{eq:simgamn}
\end{gather}
}
From the optical viewpoint, the coefficients 
$\ttr_n$ and $\ttt_n$ of $\tau$'s 
are regarded as reflection and transmission ratios respectively,
as illustrated in Figure \ref{fig:OL} (b).
The probability density of quantum walks corresponds to
the intensity of the optical wave.
The condition 
\revs{$|\ttt_{n}\ttt_{n}'-\ttr_{n}\ttr_{n}'|=1$
}
means the energy conservation.
If $\ttr_n$ and $\ttt_n$ are complex numbers,
they mean the phase shift at the interface.

The homogeneous case in which the transmission rate is
equal to the reflection rate, it corresponds  to the Hadamard
process. The peak in the distribution in Figure \ref{fig:HP}  
could be considered as
the wavefront, which speed $c$ differs from the original speed $c_0$ of
the particle but is $c= c_0/\sqrt{2}$. 
It reminds us of the fact that the group velocity in general differs
from the speed of the phase velocity.

The $A$-impurity means that the transmission ratio in a point differs
from the others whereas
$B$-impurity corresponds to the fact that
the reflection ratio in the point differs
from the others.
Thus due to the $B$-impurity, the optical lengths in the both sides
loops in Figure \ref{fig:OL} (c) 
differ from the Hadamard case and it behaves like 
the Fabry-Perot interferometer \cite{BW}. Thus $B$-impurity
causes the localization.
The light is confined there from the optical viewpoint,
though it is studied rigorously in
\cite{K10}.

\begin{figure}[ht]
 \begin{center}
\includegraphics[width=11cm]{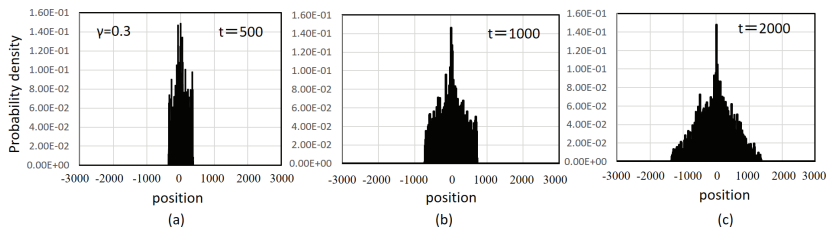}
 \newline
 \end{center}
\caption{
The intensity distribution of the optical wave of
the point  process with $\gamma =0.3$ of the seed of the
pseudo-random $=2$:
(a) $t=500$,
(b) $t=1000$ and
(c) $t=2000$.
}
\label{fig:PPtypeind}
\end{figure}

\begin{figure}[ht]
 \begin{center}
\includegraphics[width=11cm]{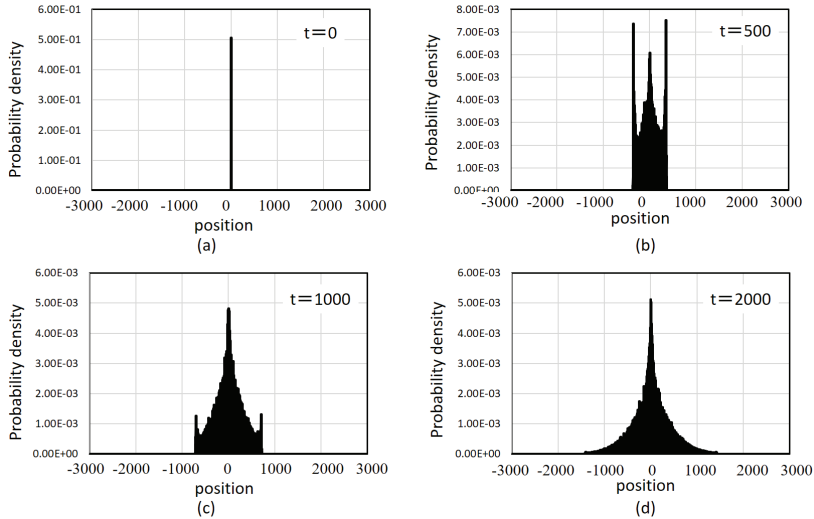}
 \newline
 \end{center}
\caption{
The average of
the intensity distribution of the optical wave of
the point  process with $\gamma =0.3$:
(a) $t=0$,
(b) $t=500$,
(c) $t=1000$ and
(d) $t=2000$.
}
\label{fig:PPtype}
\end{figure}

\begin{figure}[ht]
 \begin{center}
\includegraphics[width=11cm]{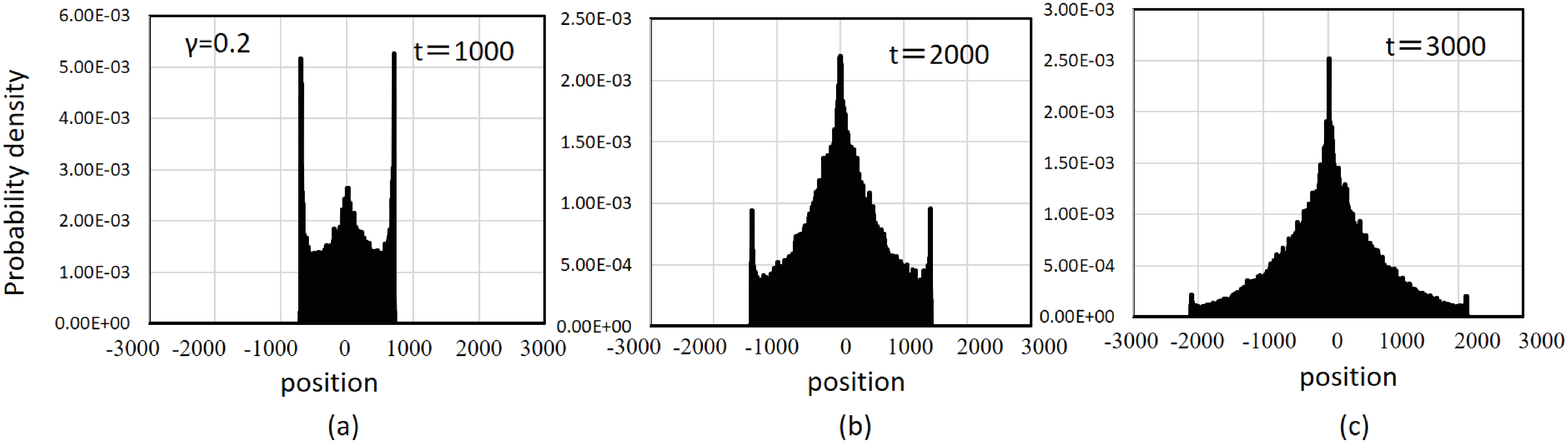}
 \newline
\includegraphics[width=11cm]{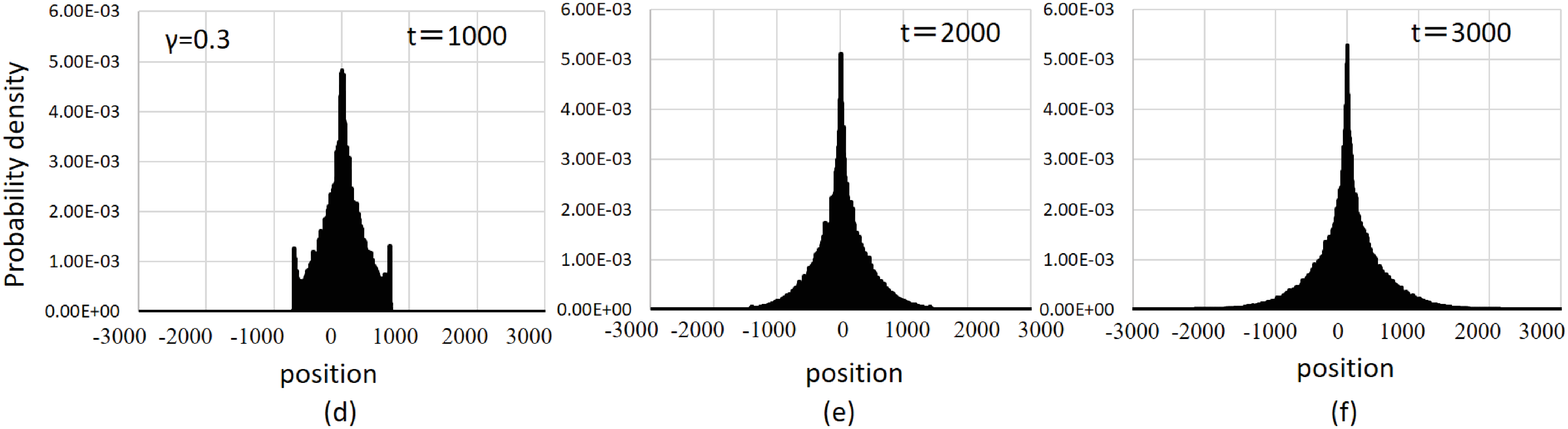}
 \newline
\includegraphics[width=11cm]{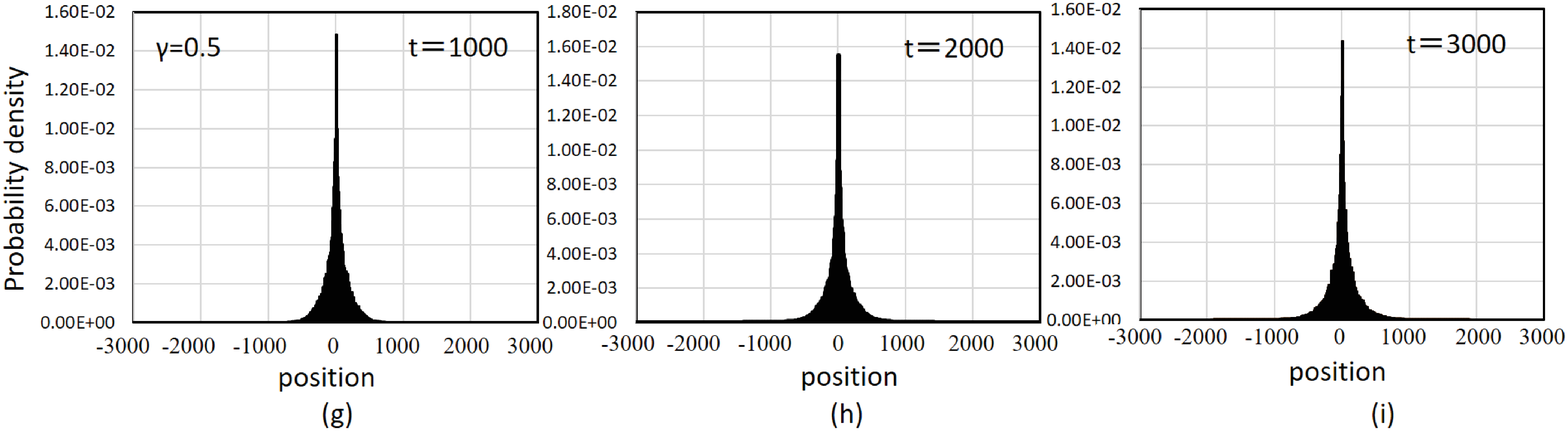}
 \newline
 \end{center}
\caption{
The average of
the intensity distributions of the optical wave of
the point  process at
\revs{
$t=1000$,  $t=2000$ and $t=3000$.
}
 (a),(b) and (c) correspond to $\gamma=0.2$,
 (d),(e) and (f) to $\gamma =0.3$, and 
 (g),(h) and (i) to $\gamma =0.5$. }
\label{fig:PPtypeII}
\end{figure}

\bigskip 
\bigskip

\begin{figure}[ht]
\begin{minipage}{0.45\hsize}
 \begin{center}
\includegraphics[width=7cm]{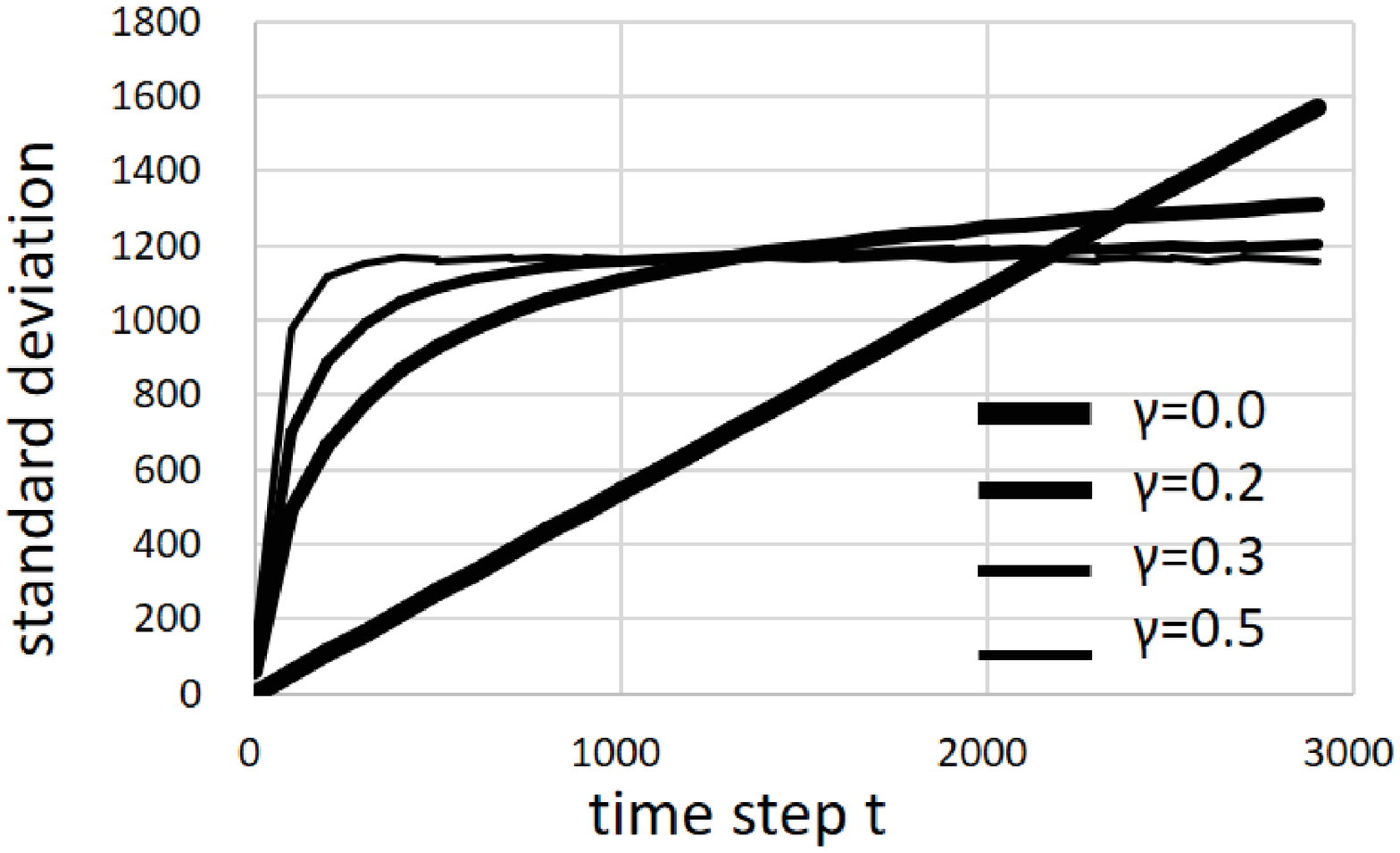}
 \newline
(a)
 \end{center}
\end{minipage}
\begin{minipage}{0.45\hsize}
 \begin{center}
\includegraphics[width=7cm]{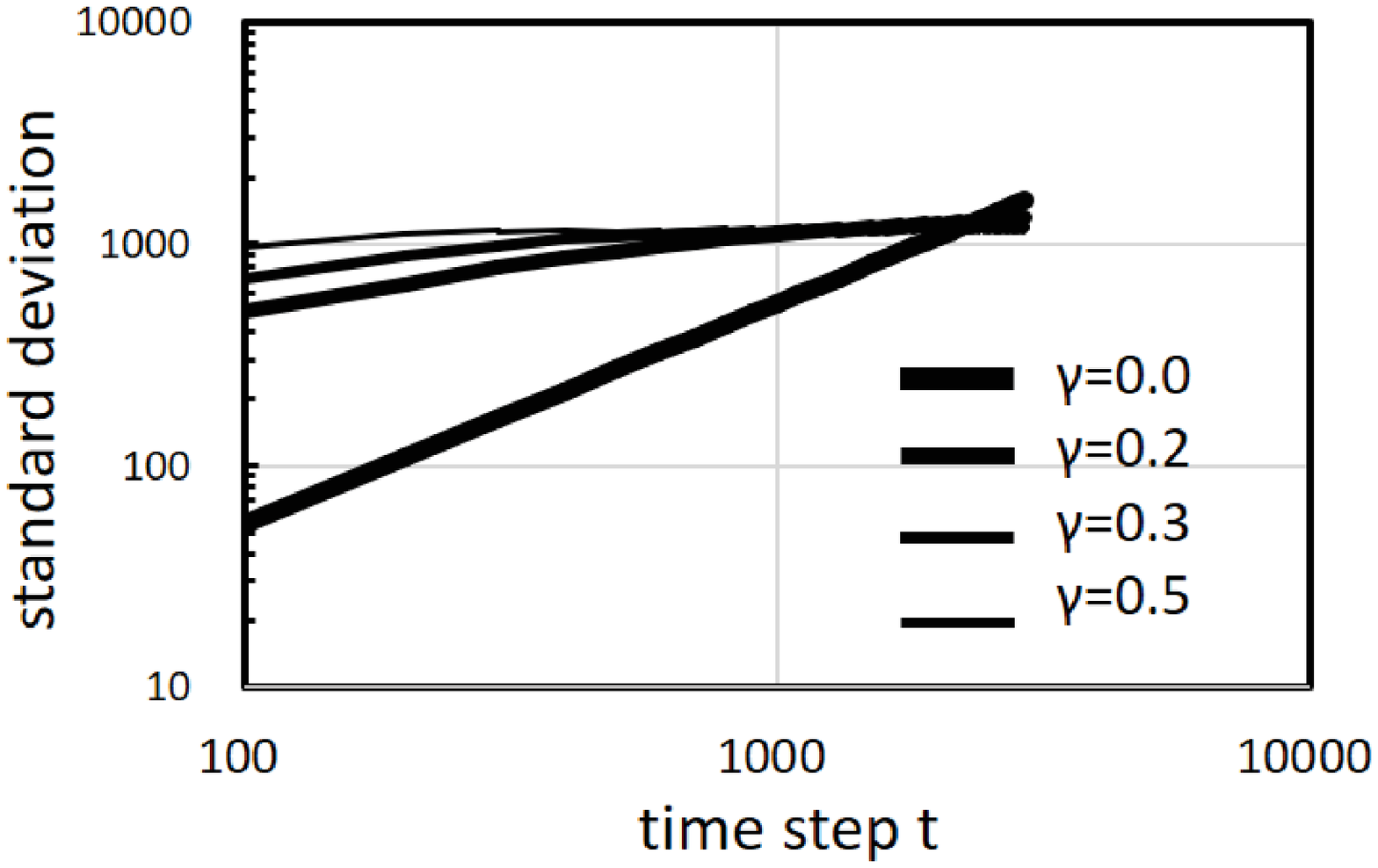}
 \newline
(b)
 \end{center}
\end{minipage}
\caption{
The time dependence of the standard deviation of
the density distribution \revs{for $\gamma = 0.0, 0.2, 0.3$ and $0.5$.}}
\label{fig:SD}
\end{figure}

\begin{figure}[ht]
\begin{minipage}{0.45\hsize}
 \begin{center}
\includegraphics[width=7cm]{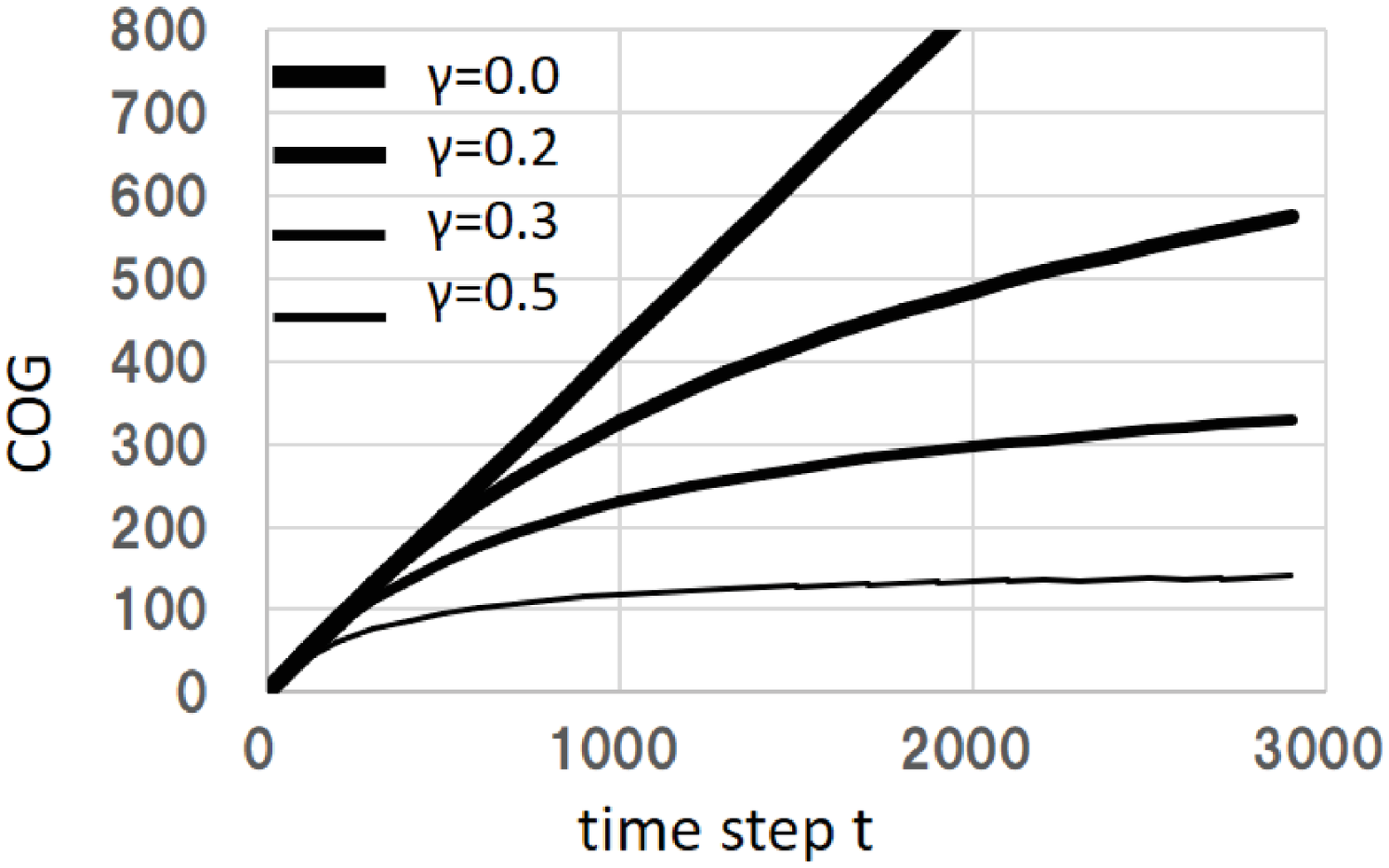}
 \newline
(a)
 \end{center}
\end{minipage}
\begin{minipage}{0.45\hsize}
 \begin{center}
\includegraphics[width=7cm]{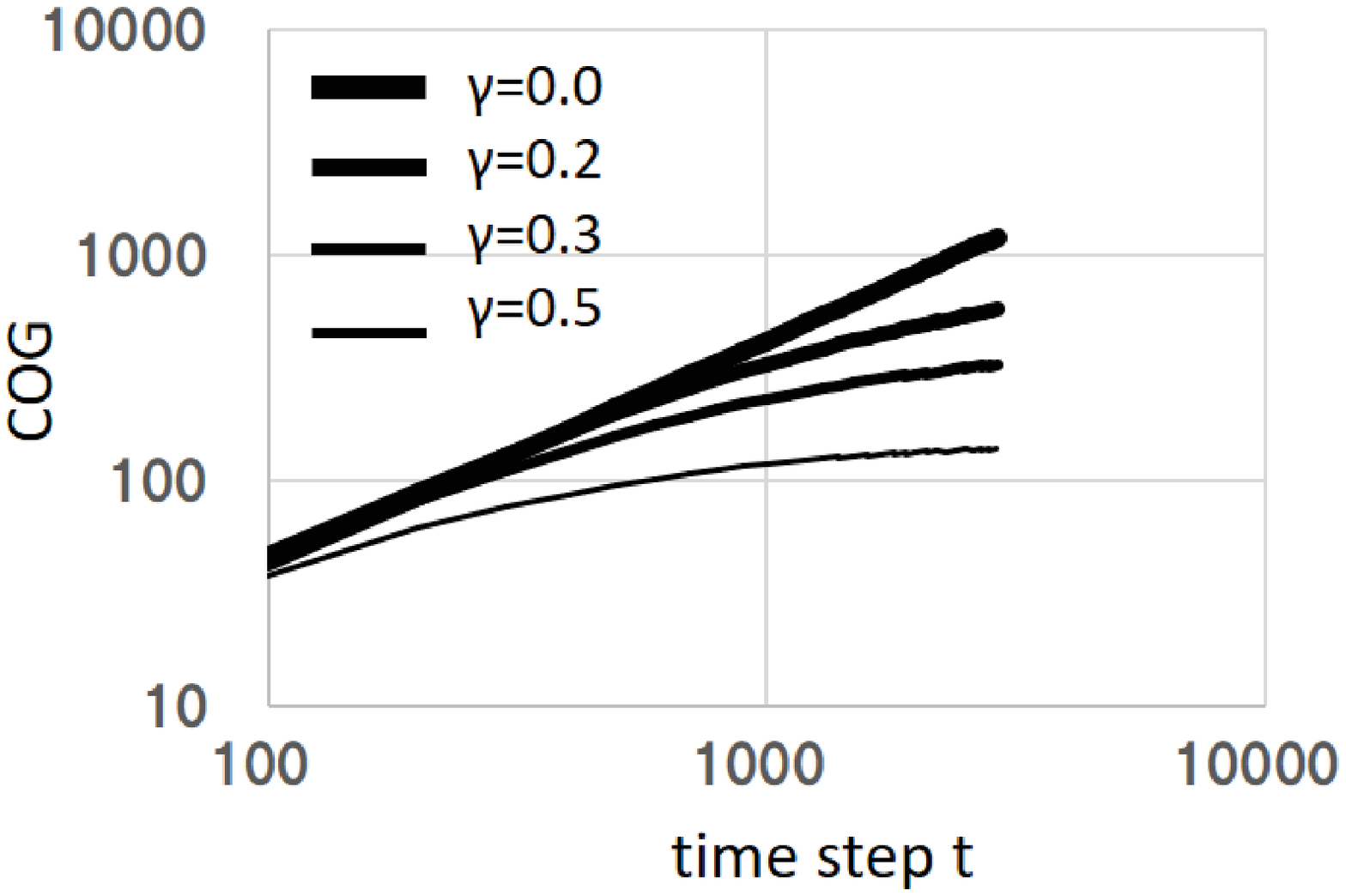}
 \newline
(b)
 \end{center}
\end{minipage}
\caption{
The time dependence of the center of gravity of the half side of
the density distribution \revs{for $\gamma = 0.0, 0.2, 0.3$ and $0.5$.}}
\label{fig:COG}
\end{figure}

\section{Localization in Poisson point process of the B-impurity}
\label{sec:PPP}

In this section, 
we consider a disordered model of quantum walks
for a random configuration of the $B$-impurities,
$$
\tau_n = 
\left\{
\begin{matrix}
\frac{1}{\sqrt{2}}
\begin{pmatrix} 1 & \ee^{\ii \gamma} \\ \ee^{-\ii \gamma} & -1 \end{pmatrix}.
& \mbox{for }n \in \cN_0,\\
\frac{1}{\sqrt{2}}
\begin{pmatrix} 1 & 1 \\ 1 & -1 \end{pmatrix},
& \mbox{otherwise}.\\
\end{matrix}\right.
$$
where the positions $\cN_0$ are given in uniformly random.
More precisely the configuration obeys the Poisson point process. 
In the numerical computations,
we used the pseudo-randomness of a certain seed $i_S$.
However if two impurities are assigned a single point, we set one of them
to the nearest empty points.
We avoid the multiple occupation of the impurities.

Then we show that such impurities localize the wave
and make it forget its wave nature.

The number of points or matrices is 6000. The number of B-impurities is
300 whose phase $\gamma$ is 0.3. The density $p$ of 
B-impurities is fixed as $0.05$.
The initial state of
every case is the same as (\ref{eq:II2}).
We computed the quantum walks numerically as in Figure \ref{fig:PPtypeind}.
We set 100 different configurations using different seeds $i_S$'s,
and computed 100 cases of the time-evolutions of
the probability densities of the quantum walks.
We averaged the densities over 100 seeds of  the pseudo-randomness
and obtained the time-evolutions of the averaged densities as
in Figure \ref{fig:PPtype}.
With the passage of time, the shape of the density differs from
that of Hadamard type as in Figure \ref{fig:HP}.
There exists a localized state around the center but it is extended 
with the passage of time.

We further handled other two cases;
the number of the points and the density of B-impurities are
the same as above but the phase $\gamma$ are 0.2 and 0.5.
The computation results of the averaged densities 
are displayed in Figure \ref{fig:PPtypeII}.
They are symmetric and show that the larger $\gamma$ is, 
the more the density is localized.
Localized densities bereave the coherency of the density.

Following the arguments in
 \cite{MBSS,RMM}, we investigated the standard deviations
of these density distributions as in Figure  \ref{fig:SD}.
However the standard deviations exhibit
 that the density distributions behave like stationary,
though the localized states change their shapes as in
Figure \ref{fig:PPtypeII}.
Since each density distribution is symmetric with respect to the origin,
we compute the time dependence of the center of gravity
of the half-side (\ref{eq:1.1})
as  which is displayed in Figure  \ref{fig:COG}.

\begin{figure}[ht]
 \begin{center}
\includegraphics[width=8cm]{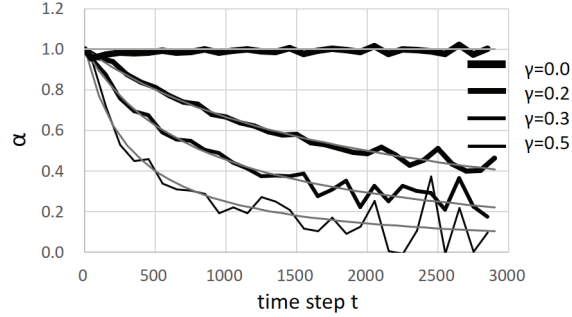}
 \end{center}
\caption{
The order $\alpha$ of
the time dependence of the center of gravity of the half side of
the density distribution.
The gray curves are given by (\ref{eq:alphat}).}
\label{fig:alphat}
\end{figure}

The behaviors of the cases $\gamma>0$ in Figure  \ref{fig:COG} 
are contrast to
the time dependence of the center of gravity of the
Hadamard process $\gamma=0$.
It is emphasized that
Figure  \ref{fig:COG} represents well the properties of
the behaviors in
Figure  \ref{fig:PPtypeII}.
The time-evolution of the center of gravity is expressed 
well by 
\begin{equation}
 \mathrm{COG}(t) = \beta t^{\alpha(t)},
\end{equation}
where $\alpha(t) \in [0,1]$.
Since $\alpha(t)$ is expressed by
$$
 {\alpha(t)} = \frac{t}{\mathrm{COG}(t)}
 \frac{d \mathrm{COG}(t)}{dt},
$$
we display $\alpha(t)$ by replacing the differential
by the finite difference procedure as in Figure  \ref{fig:alphat}.

For the formal heat kernel,
 $\frac{1}{\sqrt{4\pi t}} \ee^{x^2/4t}$, 
its center of gravity is given as
\begin{equation}
\begin{split}
\mathrm{COG}_{\mathrm{HK}}(t) 
&= \frac{\int^\infty_0 \frac{x}{\sqrt{4\pi t}} \ee^{-x^2/4t} dx}
        {\int^\infty_0 \frac{1}{\sqrt{4\pi t}} \ee^{-x^2/4t} dx}\\
&= \frac{\int^\infty_0 \sqrt{t}\frac{s}{\sqrt{2\pi}} \ee^{-s^2} ds} 
        {\int^\infty_0 \frac{1}{\sqrt{2\pi}} \ee^{-s^2} ds}\\
&= \sqrt{\frac{t}{2\pi}}.
\end{split}
\end{equation}
and thus the diffusion process has
the parameter $\alpha =\alpha_{HK}=\frac{1}{2}$ of 
COG.

In the plots in Figure  \ref{fig:alphat}, 
we can approximate them by the formula
\begin{equation}
   \alpha(t) = \frac{1}{\kappa t+1}.
\label{eq:alphat}
\end{equation}
In Figure  \ref{fig:alphat}, 
gray curves are given by the
equation (\ref{eq:alphat}): $\kappa = 0.05$ for $\gamma=0.2$ case,
$\kappa = 0.0012$ for $\gamma=0.3$ case.
and $\kappa = 0.003$ for $\gamma=0.5$ case.
The numerical computations show that they vanish asymptotically.
Since in Figure \ref{fig:COG}, 
$\alpha(t)$ of $\gamma=0.2$ has 
 the region whose the exponent $\alpha=1/2=\alpha_{HK}$,
the light behaves like diffusion process in the region.
However it is a transient. In other words,
due to the impurities, the transport states changes from
the coherent one to the localized one and in its transient,
the state like the diffusion appears.

It means that our model shows
 the coexistence
of the diffusion light and the coherence light.
With progress of time, the ratio of the coexistence changes.

\begin{figure}[ht]
 \begin{center}
\includegraphics[width=11cm]{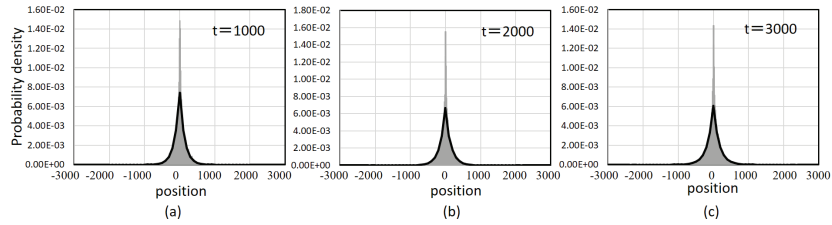}
 \end{center}
\caption{
The average of
the intensity distributions of the optical wave of
the point  process with $\gamma = 0.5$ at
\revs{
(a) $t=1000$,  (b) $t=2000$ and (c) $t=3000$.
}
}
\label{fig:PPtypeLap}
\end{figure}

Further Figure  \ref{fig:PPtypeLap} shows that
the probability density or the intensity is expressed well by 
the Laplace distribution,
\begin{equation}
P(x)dx = \frac{A}{2\delta_t}\exp\left(-
\left|\frac{x-x_0}{\delta_t}\right|\right)d x.
\end{equation}
It might be contrast to the fact that the quantum random walk
is expressed by the binomial distribution.
Behind it, we have 
$$
\left|\frac{d}{dx}P \right|
= -\frac{1}{\delta_t}P.
$$
It is decay by $-1/\delta_t$, which is obtained in \cite[(A.8)]{HvW}.

\begin{figure}[ht]
 \begin{center}
\includegraphics[width=7cm]{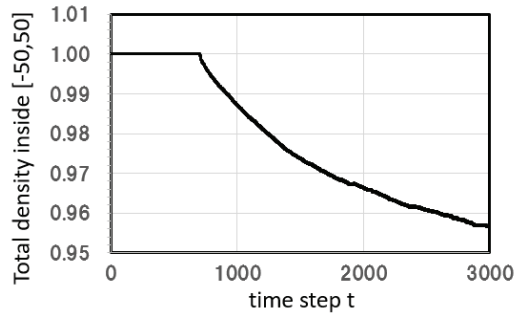}
 \end{center}
\caption{
Time dependence of the total density inside $[-50,50]$ for 
the point  process $\gamma = 0.5$} 
\label{fig:PPtypePen}
\end{figure}

\revs{
However even
}
 for the $\gamma=0.5$ case, the penetration to outside does not
finish because the total density inside $[-50,50]$ decreases as in 
Figure  \ref{fig:PPtypePen}.
\ins{
In fact, (\ref{eq:alphat}) shows that the behavior of the intensity
 becomes stationary following the power law.
It means that
}
the localized state changes slowly.

\bigskip
\ins{
In \cite{G},
Goedecke assumed that due to the randomness, the
correlation
$\langle \sigma_{n}^{(+)} \overline{\sigma_{n}^{(-)}}\rangle$ 
vanishes and derived the Kubelka-Munk equation.
Here 
$\langle \sigma_{n}^{(+)} \overline{\sigma_{n}^{(-)}}\rangle$ 
is the correlation averaged over the space direction.
Further in \cite{HvW}, under the same assumption,
Haney and van Wijk derived the time development equation
which they called the modified Kubelka-Munk equation.
}

\begin{figure}[ht]
 \begin{center}
\includegraphics[width=7cm]{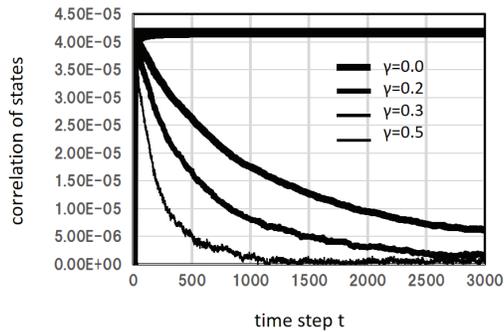}
 \end{center}
\caption{
The time-dependence of the real part of the 
correlation $\eta$ for 
$\gamma = 0.0, 0.2, 0.3$ and $0.5$.
}
\label{fig:eta}
\end{figure}

\ins{
Thus we also compute the correlation but due to
symmetry of this system, the correlation always
vanishes.
Hence we computed the time-dependence of 
the real part of the correlation quantity of the half-side,
\begin{equation}
\eta:= Re\sum_{n \ge 0}\sigma_{n}^{(+)}
\overline{\sigma_{n}^{(-)}} ,
\label{eq:eta}
\end{equation}
which is illustrated in Figure \ref{fig:eta}.
Whereas the real part $\eta$ of
$\langle \sigma_{n}^{(+)} \overline{\sigma_{n}^{(-)}}\rangle$ 
is constant for $\gamma=0$ case, 
it decreases with the passage of time for $\gamma \neq 0$ case.
Further the larger $\gamma$ is, the correlation
$\eta$ decreases more rapidly.
The insight of Goedecke in \cite{G} means that
these systems in this article change from the coherent one to
diffusive one.
}

\section{Discussion}
\label{sec:Dis}

The time dependence of the distribution in the Poisson point process
of B-impurities is
quite different from the pure Hadamard process.
It shows that the B-impurities
suppress the coherent properties whereas
the pure Hadamard process shows the coherent nature.

Further in the intermediate process, the ray is decomposed to
the localized one and the coherent one. It means the
coexistence of the coherent state and diffusive state.
Thus it implies that
the behavior of the light ray in the real material could be
approximated well by the coherent light governed by Fermat
principle and diffusive light.

For example, in the real material, there are so many impurities
and thus, these numerical computations support the coexistence
of two facts that 1)
the coloring is predicted well by Kubelka-Munk formula \cite{D},
in which the coherence of the wave is neglected, 
and 2) the light is expressed well by the coherent wave as in \cite{F}.

\ins{
The obtained results are very similar to the numerical
computational results in \cite{HvW}. 
They studied the same problem based on the wave equation.  
However since it is not clear the relation between
the continuum space-time version of  quantum walks and wave equation, 
in this stage, we cannot give the further investigation 
on the relation to the results in \cite{HvW} rigorously.
}

\ins{
However in the framework of the path integral method and 
transfer matrix theory, we could give comments on our results.
Our model in quantum walks is naturally obtained
as the extension of the path integral and transfer matrix theory
as in \ref{sec:TMPIQW}. 
We computed the correlation (\ref{eq:eta})
of the wave function as in Figure \ref{fig:eta};
Goedecke derived the Kubelka-Munk equation from the 
transfer matrix theory by assuming
the fact that the randomness makes the correlation
$\langle \sigma_{n}^{(+)} \overline{\sigma_{n}^{(-)}}\rangle$ vanish
\cite{G}.
Though we do not give the direct relation to these investigations,
our computations in our model of quantum walks also provide
the justification of his assumption.
}

\ins{
The advantage of our model in quantum walks is to give
rigorous arguments without problems of discretization as 
mentioned in Introduction.
It is  simple and rigorous though we did not give their explicit
relations to  the wave equation as the continuum limit.
Thus it should be regarded as a toy model but
its connectivity with the transfer matrix theory and path integral
is natural. Thus we consider that this model gives a new
theory in optics to analyze several fundamental problems including this
decoherent problem as quantum walks provide rigorous reinvestigations
on crucial physical phenomena.
}

Since it is known that the \revs{structural}
 color is caused by the interference
of the light,
if we unify the coloring in the diffusive reflection  and
the \revs{structural} color, our theory has an effect on such investigation.

Further we should note that
our investigation might be related to the random lasers
\cite{CLFAR}.

\section*{Acknowledgments}
This work is dedicated to Professor Kenichi Tamano.
The authors would like to thank him
for encouragements, suggestions and profound discussions.
This work has been done in his seminar which started 1988.

Y. I. was supported by the Grant-in-Aid for Young Scientists (B)
 of Japan Society for the Promotion of Science (Grant No. 16K17652). 
N. K. was supported by the Grant-in-Aid for Challenging Exploratory 
Research of Japan Society for the Promotion of 
Science (Grant No. 15K13443). 
S. M. was supported by the Grant-in-Aid for Scientific Research (C) 
of Japan Society for the Promotion of Science (Grant No. 16K05187). 
H. M. was supported by the Grant-in-Aid for Scientific Research (C) 
of Japan Society for the Promotion of Science (Grant No. 16K05249). 

\ins{
The authors are grateful to the referees for critical and
helpful suggestions, especially for pointing out the 
reference \cite{HvW}.
}

\appendix
\section{Transfer matrix, S-matrix and quantum walk}
\label{sec:TMPIQW}

\ins{
In this appendix, we review the transfer matrix theory with
$S$-matrix and the path integral (path summation)
method, and show their connection with quantum walks.
}

\begin{figure}[ht]
\begin{minipage}{0.45\hsize}
 \begin{center}
\includegraphics[width=5cm]{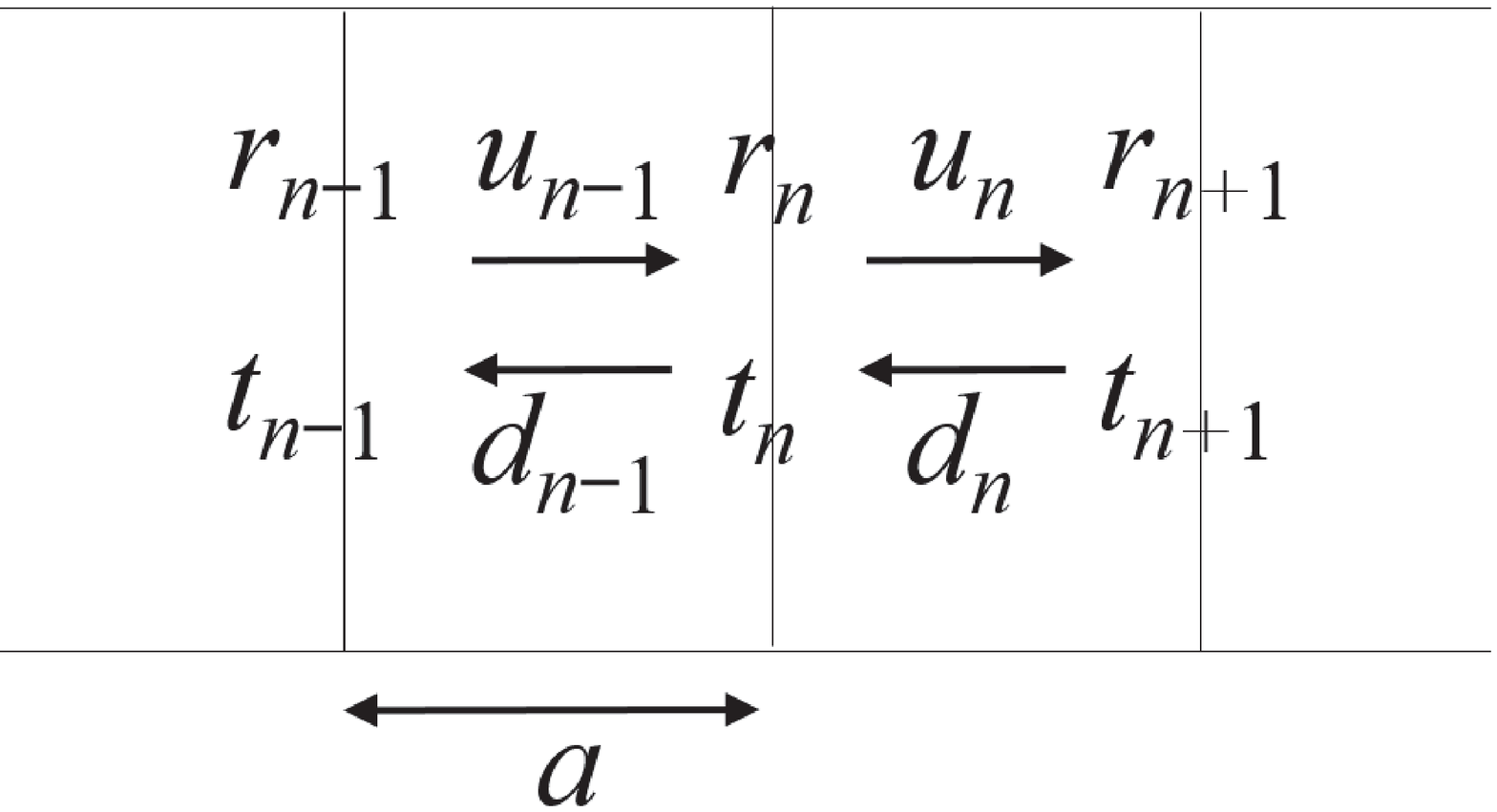}
 \newline
(a)
 \end{center}
\end{minipage}
\begin{minipage}{0.45\hsize}
 \begin{center}
\includegraphics[width=4cm]{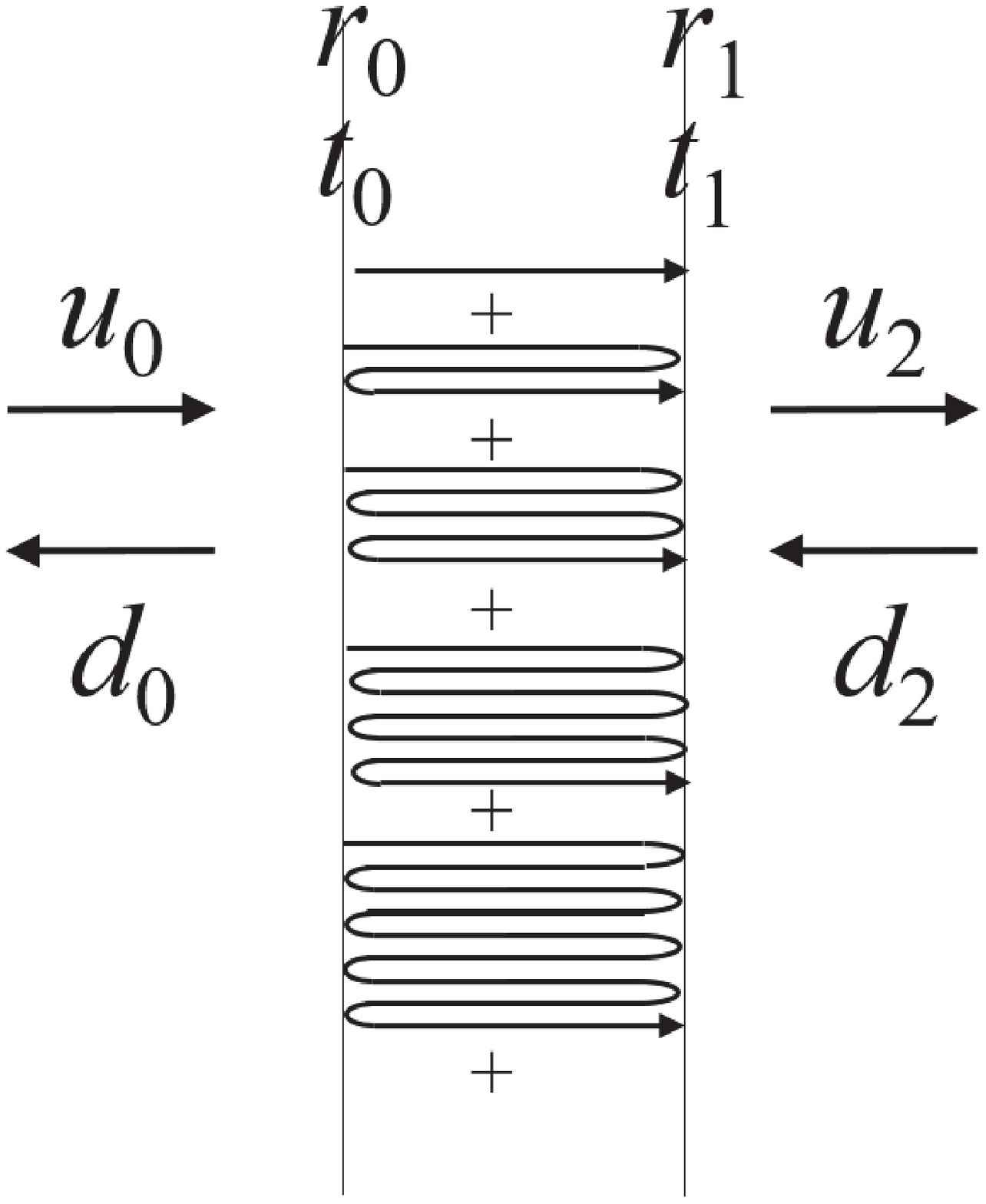}
 \newline
(b)
 \end{center}
\end{minipage}
\caption{
Optical system}
\label{fig:TMPI}
\end{figure}

\ins{
We consider the scalar Helmholtz equation, which is also
the fundamental equation of
the Kirchhoff wave optics \cite{BW}. 
In this article, we deal with its one-dimensional
case,
\begin{equation}
\frac{d^2}{d z^2} \psi +\frac{\omega^2\mathfrak{n}(x)}{c^2} \psi = 0,
\label{eq:Keq}
\end{equation}
where $\mathfrak{n}$ is the optical index and $\omega$ is the frequency.
As we displayed it in Figure \ref{fig:TMPI} (a),
in order to solve this equation, we divide the one-dimensional region
to $N$ pieces with the interval $a$ , $[n a, (n+1)a)$, $(n=0, 1, 2, \cdots, N)$
and assume that $\mathfrak{n}$ is constant at each $[n a, (n+1)a)$.
In order to solve (\ref{eq:Keq}), we set
$$
\psi(x) = u_n \ee^{\ii k_n (x-n a)}
+ d_n \ee^{-\ii k_n (x-a n)} , \quad
x \in [n a,  (n+1)a),
$$
where $k_n^2 = \omega^2\mathfrak{n}(x)/c^2$.
At each $x \in 2an$, $(n=0, 1, 2, \cdots, N)$,
the solution of the Helmholtz equation must
satisfy
$$
    \psi(na)
    =\lim_{x \to n a -0} \psi(x), \quad
    \left(\frac{d}{dx}\psi\right)(na)
    =\lim_{x \to n a-0}\frac{d}{dx} \psi(x). \quad
$$
These mean
\begin{equation}
\begin{pmatrix}
1 &
1 \\
k_{n}& -k_{n} \\
\end{pmatrix}
\begin{pmatrix}
u_{n}\\
d_{n}
\end{pmatrix}
=
\begin{pmatrix}
\alpha_{n-1} &
1/\alpha_{n-1} \\
k_{n-1}\alpha_{n-1} &
-k_{n-1}/\alpha_{n-1} \\
\end{pmatrix}
\begin{pmatrix}
u_{n-1}\\
d_{n-1}
\end{pmatrix},
\label{eq:TM1}
\end{equation}
where $\alpha_n = \ee^{\ii k_n a}$.
By letting 
$$
\hht_n:=\frac{2k_n}{(k_n + k_{n-1})},\quad
\hhr_n:=\frac{(k_n - k_{n-1})}{k_n+k_{n-1}},\quad
$$
(\ref{eq:TM1}) is reduced to
$$
\begin{pmatrix}
u_{n}\\
d_{n}
\end{pmatrix}
=
T^\alpha_n
\begin{pmatrix}
u_{n-1}\\
d_{n-1}
\end{pmatrix} 
=
T_n
\begin{pmatrix}
\alpha_{n-1} & 0\\
0& 1/\alpha_{n-1}
\end{pmatrix} 
\begin{pmatrix}
u_{n-1}\\
d_{n-1}
\end{pmatrix}, 
$$
where 
\begin{gather*}
\begin{split}
T_n :=
\begin{pmatrix}
T_{n,11} & T_{n,12} \\ 
T_{n,21} & T_{n,22}  
\end{pmatrix}
&=
\frac{1}{2k_n}
\begin{pmatrix}
(k_n + k_{n-1}) & (k_n - k_{n-1}) \\ 
(k_n - k_{n-1}) & (k_n + k_{n-1})  
\end{pmatrix}\\
&=
\frac{1}{\hht_n}
\begin{pmatrix}
1 & \hhr_n \\
\hhr_n  & 1\\
\end{pmatrix}.\\
\end{split}
\end{gather*}
On the other hand, when we are concerned with 
the relations between incoming rays, $u_{n-1}$, $d_n$ and
out-going rays, $u_n$, $d_{n-1}$, we investigate
$S$-matrix, $S_n$, at each point $x=na$,
\begin{equation}
\begin{pmatrix}
u_{n}\\
d_{n-1}/\alpha_{n-1}
\end{pmatrix} = S_n
\begin{pmatrix}
u_{n-1}\alpha_{n-1}\\
d_{n}
\end{pmatrix}, \quad 
S_n =
\begin{pmatrix}
t_n & r_n \\
r_n' & t_n' 
\end{pmatrix} =
\frac{1}{T_{n,22}}
\begin{pmatrix}
\det(T_{n}) & T_{n,12}\\
-T_{n,21} & 1
\end{pmatrix}
\label{eq:S_n}
\end{equation}
Here 
$r_n$'s and $t_n$'s are the reflection and the transmission ratios
respectively. 
In this case, $r_n=-r_n'=\hhr_n$, $t_n'= \hht_n$ and 
$t_n=\displaystyle{\frac{k_{n-1}}{k_n}\hht_n=
\frac{2 k_{n-1}}{k_n+k_{n-1}}}$.
It is turned out that 
$S_n$ is an unitary matrix and 
the total intensity of the incoming rays is equal to 
that of the outgoing rays, i.e.,
$|u_{n}|^2 + |d_{n-1}|^2 = |u_{n-1}|^2 + |d_{n}|^2$.
Using the transfer matrices 
${\mathbf{T}}_{0,N}=T^\alpha_{N}T^\alpha_{N-1}\cdots 
T^\alpha_2T_1$,
we also have
$$
\begin{pmatrix}
u_{N}\\
d_{0}/\alpha_0
\end{pmatrix} = {{\mathbf{S}}}_{0,N}
\begin{pmatrix}
u_{0}\alpha_0\\
d_{N}
\end{pmatrix}, \quad 
$$
which is obtained like $S_n$ for $T_n$.
For example, $N=2$ case illustrated in Figure \ref{fig:TMPI} (b),
\begin{equation}
{{\mathbf{S}}}_{0,1}=
\begin{pmatrix}
t_2(1+r_1 r_2\alpha_2^2)^{-1} t_1 &
r_2+t_2t_2'(1+r_1 r_2\alpha_2^2)^{-1}\\
r_1+t_1t_1'(1+r_1 r_2\alpha_2^2)^{-1}  &
t_2(1+r_1 r_2\alpha_2^2)^{-1} t_1
\end{pmatrix}.
\label{eq:S0N}
\end{equation}
Here we use the facts that $r_n^2 + t_n t_n' = 1$ and $r_n' = - r_n$.
}

\ins{
In the path integral method, we sum the amplitude over all
possible paths as in \cite{C} noting that
there is the phase difference between the reflection ratios, 
i.e., $r'_n=-r_n$. The amplitude is determined by the 
multiplication of $r_n$'s, $t_n$'s and the phase factor
$\ee^{\ii k_n L}$, where $L$ is the length of the path.
For example, for the simple case as in Figure \ref{fig:TMPI}, 
we sum the amplitudes over all possible paths
and by using the relation, 
$$
 (1-r_1r_2\alpha_2^2+(r_1r_2 \alpha_2^2)^2-\cdots) = 
\frac{1}{1+r_1r_2 \alpha_2^2},
$$
we recover (\ref{eq:S0N}) even for the path integral (path
summation) method.
Crook proved that
the results of the path integral method are completely identified with 
those of the $S$-matrix method for more general cases \cite{C}.
}

\ins{
However these are static.
In quantum walks, we introduce the
time parameter $m$.
Using the fact that
$|u_{n}|^2 + |d_{n-1}|^2 = |u_{n-1}|^2 + |d_{n}|^2$,
let $\displaystyle{
\begin{pmatrix}
\sigma_{n,m}^{(+)}\\
\sigma_{n,m}^{(-)}\\
\end{pmatrix}
=\begin{pmatrix}
u_{n-1} \alpha_{n-1}\\
d_{n}
\end{pmatrix}
}$.
Noting the relations,
$$
\begin{pmatrix}
u_{n-1}\alpha_{n-1}\\ 0
\end{pmatrix}
=
\begin{pmatrix}
 \alpha_{n-1}t_{n-1} & \alpha_{n-1} r_{n-1}\\
0 & 0
\end{pmatrix}
\begin{pmatrix}
u_{n-2} \alpha_{n-2}\\
d_{n-1}
\end{pmatrix}, \quad
$$
$$
\begin{pmatrix}
0 \\ d_{n}
\end{pmatrix}
=
\begin{pmatrix}
0 & 0\\
\alpha_{n} r_{n+1}' & \alpha_n t'_{n+1}\\
\end{pmatrix}
\begin{pmatrix}
u_{n}\alpha_{n}\\
d_{n+1}
\end{pmatrix}
$$
we obtain the time-evolution (\ref{eq:simgamn}) by letting
$\ttt_n = \alpha_n t_n$, $\ttt_n' = \alpha_{n-1} t_n$
$\ttr_n = \alpha_n r_n$ and
$\ttr_n' = \alpha_{n-1} r_n'$. It is obvious that
the associated matrix is unitary.
}

\section{Short Review of Kubelka-Munk Theory}
\label{sec:KMT}

In \cite{KM}, Kubelka and Munk investigated the diffusive light whose
value is given by real number in paint layer.
They considered the one-dimensional half space whose negative is for downward
region which consists of paint layer. They modeled the diffusive light
by  the downward-going light $i$ and  the upward-going light
$j$ which interchanges by scattering and absorption for
the scattering ration $s$ and an absorption ration $k$:
$$
\frac{d}{dx} 
\begin{pmatrix} i\\ j \end{pmatrix}
=\begin{pmatrix} s + k & -s \\ s & -(s+k) \end{pmatrix}
\begin{pmatrix} i\\ j \end{pmatrix}.
$$
Since this matrix denoted by $S$ has the property,
$$
S^2 = 
\begin{pmatrix} (s + k)^2-s^2 & 0\\ 0 & (s+k)^2 - s^2 \end{pmatrix},
$$
we have 
$$
\ee^{S x} = \cosh(\sqrt{(s+k)^2-s^2}x) I
 +\frac{\sinh(\sqrt{(s+k)^2-s^2}x)}{ 
 \sqrt{(s+k)^2-s^2}} S,
$$
and
$$
\begin{pmatrix} i(x)\\ j(x) \end{pmatrix}
=\exp(S x) 
\begin{pmatrix} i_0\\ j_0 \end{pmatrix},
$$  
where $I$ is the unit matrix and $i_0$ and $j_0$ are
the intensity of each light at $x=0$.

Noting asymptotically 
$$
\cosh(\sqrt{(s+k)^2-s^2}x)/\sinh(\sqrt{(s+k)^2-s^2}x) \to -1
$$ for $x\to-\infty$,
if we assume, for example, that $i(x)$ vanishes
at $x\to-\infty$, we have 
$$
j_0=\left(1+\frac{k}{s}-\sqrt{\frac{k^2}{s^2}+\frac{2k}{s}}\right)i_0.
$$
For the case, the ratio of the intensities of
the diffusive light and the incoming ray,  
$$
R_\infty=\frac{j_0}{i_0},
$$
is expressed by
$$
R_\infty=\left(1+\frac{k}{s}-\sqrt{
\frac{k^2}{s^2}+\frac{2k}{s}}\right), \quad
\frac{k}{s}=\frac{(1-R_\infty)^2}{2R_\infty}.
$$
By observing the reflection ratio
$R_\infty$ for an ideal case for monochromatic light with
the wave length $\lambda$,
 we determine the ratio $k/s$ each $\lambda$. Using $k/s$, 
we can predict the coloring for the light.
For example, if we use $n$ paints $(k_i, s_i)$ in layers with
depth $d_i$,
we have the formula
$$
\begin{pmatrix} i(\sum_id_i)\\ j(\sum_i d_i) \end{pmatrix}
= \ee^{S_n d_n} \cdots \ee^{S_1 d_1}
\begin{pmatrix} i_0\\ j_0 \end{pmatrix},
$$
where $S_i$ is the matrix $S$ consisting of $(k_i, s_i)$.
if 
$\begin{pmatrix} i(\sum_id_i)\\ j(\sum_i d_i) \end{pmatrix}$ is 
determined by a boundary condition,
we have
$$
\begin{pmatrix} i_0\\ j_0 \end{pmatrix}
= \ee^{-S_1 d_1} \cdots \ee^{-S_n d_n}
\begin{pmatrix} i(\sum_id_i)\\ j(\sum_i d_i) \end{pmatrix}.
$$
This shows the reflection ratio of the light for layers.
This theory is a nice theory to predict the coloring in nature and
has been used in painting and printing industry including cosmetics for
protection of the ultraviolet rays.

However these rays are given by real numbers whereas the light is basically 
given by the complex number as in \cite{F}. 
The difference are concerned.

\end{document}